\documentclass[10pt, onecolumn]{article} 

\usepackage[english]{babel}
\usepackage[utf8]{inputenc}
\usepackage[onehalfspacing]{setspace}
\usepackage[margin=25mm]{geometry}

\usepackage{graphicx}
\usepackage{url,hyperref,lineno,microtype,subcaption}
\usepackage[onehalfspacing]{setspace}
\usepackage{algorithm, xcolor, tabularx}
\usepackage{threeparttable}
\usepackage{tablefootnote}
\usepackage{authblk}

\begin{document}

\title {Neural Network Methods for Radiation Detectors and Imaging}

\author[1,2]{S. Lin}
\author[2]{S. Ning}
\author[2]{H. Zhu}
\author[3]{T. Zhou}
\author[1]{C. L. Morris}
\author[1]{S. Clayton}
\author[4]{M. Cherukara}
\author[2,5,6,*]{R. T. Chen}
\author[1,*]{Z. Wang}

\affil[1]{Los Alamos National Laboratory, Los Alamos, NM 87545, USA}
\affil[2]{Department of Electrical and Computer Engineering, The University of Texas at Austin, Austin , TX 78705, USA}
\affil[3]{Center for Nanoscale Materials, Argonne National Laboratory, Lemont , IL 60439, USA}
\affil[4]{Advanced Photon Source, Argonne National Laboratory, Lemont , IL 60439, USA}
\affil[5]{Microelectronics Research Center, The University of Texas at Austin, Austin , TX 78758, USA}
\affil[6]{Omega Optics, Inc., Austin , TX 78757, USA }
\affil[*]{Correspondence: R. T. C. (chenrt@austin.utexas.edu), Z. W. (zwang@lanl.gov) }

\date{}

\maketitle

\begin{abstract}
Recent advances in image data processing through machine learning and especially deep neural networks (DNNs) allow for new optimization and performance-enhancement schemes for radiation detectors and imaging hardware through data-endowed artificial intelligence. We give an overview of data generation at photon sources, deep learning-based methods for image processing tasks, and hardware solutions for deep learning acceleration. Most existing deep learning approaches are trained offline, typically using large amounts of computational resources. However, once trained, DNNs can achieve fast inference speeds and can be deployed to edge devices. A new trend is edge computing with less energy consumption (hundreds of watts or less) and real-time analysis potential. While popularly used for edge computing, electronic-based hardware accelerators ranging from general purpose processors such as central processing units (CPUs) to application-specific integrated circuits (ASICs) are constantly reaching performance limits in latency, energy consumption, and other physical constraints. These limits give rise to next-generation analog neuromorhpic hardware platforms, such as optical neural networks (ONNs), for high parallel, low latency, and low energy computing to boost deep learning acceleration.

\noindent (LA-UR-23-32395)

\tiny
\end{abstract}


\newpage

\section{Introduction}
X-rays produced by synchrotrons and free electron lasers (XFELs), together with high-energy photons above 100 keV, which are often generated using high-current (kA) electron accelerators and lately high-power lasers, are  widely used as radiographic imaging and tomography (RadIT) tools to examine material properties and their temporal evolution. Spatial resolution ($\delta$) down to atomic dimensions is possible by using diffraction-limited X-rays, $\delta \sim \lambda/2$, corresponding to Abbe's diffraction limit for X-ray wavelength $\lambda$. The overall object size that X-rays can probe readily reaches a length ($L$) greater than 1 mm, which is limited by the X-ray attenuation length and is X-ray energy dependent. In room-temperature water,  for example, $L =$ 0.19, 1.2, 5.9, and 14.1 cm for $1/e$-attenuation length of 10 keV, 20 keV, 100 keV, and 1 MeV X-rays, respectively. The temporal resolution has now approached a few femtoseconds by using XFELs, where an XFEL experiment can be repeated for many hours in a pump-probe configuration. In other words, the spatial dynamic range (i.e. for 10 keV X-rays, L $\sim$ 1 mm) is $2L/\lambda > 10^7$ and temporal dynamic range is $>$ 10$^{18}$.  Such ultra-wide-dynamic-range abilities of X-ray and photon techniques to connect elementary atomic and molecular processes, which are described by quantum physics and happen ultra fast (sub-nanosecond), with emergent macroscopic material properties and functions, which are usually treated classically through continuum approximations, make them extremely valuable in a wide range of applications, such as medicine (i.e. new drug discovery), high-energy density battery development, and applications in materials exposed to high-temperature, high radiation, and other harsh or `extreme' conditions. Additional applications include the optimization of chemical catalysis and the development of new superconductors and other quantum materials for information technology, accelerated computing, and artificial intelligence (AI).

The enormous spatial and temporal dynamic ranges give rise to `big data' in X-ray imaging, tomography, and photon science. Theoretically, 1 mm$^3$ of water contains about 5.6 $\times$ 10$^{-5}$ mole of water molecules ($N$ = 3.3 $\times$ 10$^{19}$). If the position of every molecule were recorded, the memory size would be $N\log_2N$ ($\log_2 N$ is the bit length for a binary data system) or 2.2 $\times$ 10$^{21}$ bits. In experiments, explosive data growth in X-ray and other forms of RadIT is built upon steady progress for more than 120 years in X-ray and radiation sources, detectors, computation, and lately data science. The fourth generation synchrotrons such as APS-U \cite{DBB:2022} and PETRA IV~\cite{SAB:2018} will have a significant reduction in emittance and a brilliance increase by a factor about 10$^3$ over the parameters of the third generation synchrotrons such as APS and PETRA III.  XFELs, which are many of orders of magnitude brighter than synchrotrons, will run at a higher repetition rate up to 1 MHz \cite{HDL:2021}. LCLS, in comparison, operates at 120 Hz. High-speed detectors with frame rate frequencies 
above 1 MHz are commercially available. The combination of high-repetition-rate experiments with a mega-pixel and larger recording system leads to high data rates, exceeding 1 TB/s (1 TB = 10$^{12}$ bytes), as we discuss further in Sec.~\ref{sec:det}.

Big data not only presents a significant challenge to data handling in terms of computing speed, computing power, short- and long-term computer memory, and computer energy consumption, which all together is called `computational resources', but also offer a transformative approach to process and interpret data, {\it i.e.} machine learning (ML) and artificial intelligence (AI) through data-enabled algorithms. Such algorithms, including deep learning (DL) \cite{lecun2015deep,Goodfellow-et-al-2016}, are distinctive from traditional physics, statistical, and other forward-model- or domain-knowledge-driven algorithms. Traditional algorithms are based on the domain knowledge, such as physics and statistics, and applicable to both small or large ensembles of data. In contrast, data-driven models may only rely on data explicitly for model training (tuning), model validation and use, with no domain knowledge required. In practice, domain knowledge always helps, partly due to the fact that some aspects of data models, such as the model architecture and other hyper-parameters, are chosen pragmatically and do not depend on the data. The amount of data required for data model training depends on the number of model parameters such as weights, activation functions, the number of nodes, {\it etc.} It is not uncommon that a deep neural network (DNN) may contain billions of tunable free parameters, which require a commensurate amount of data for training. Hybrid approaches to ML and AI, which merge data and domain knowledge, are increasingly popular.  Hybrid models not only supplement data-driven models with domain knowledge and reduce the amount of data required for training, but also accelerate the computational speed of traditional forward models by 10 to more than 100 times by bypassing some detailed and time-consuming computations.

We may differentiate two approaches to ML and AI by the computational resources involved and how the resources are distributed. In the {\it centralized} approach, data are collected from distributed locations or different data acquisition instruments through the internet. The data are then stored in a data center, and processed by high-performance computers or mainframes. Very large traditional, ML, or hybrid algorithms can be deployed in the data center, which also requires correspondingly large memory, energy and power consumption. Estimated global data center electricity consumption in 2022 was 240 - 340 TWh \cite{MSL:2020,datacenter}, or around 1-1.3\% of global final electricity demand. Cloud computing and data centers are now widely used to process `big data' in industry, health care, and research institutions. Through the cloud computing and data center approach, data generation and data processing tasks can be separated, which can mitigate the computation and data processing burden on people who generate data. In the {\it distributed} or {\it edge} approach, ML and AI, together with the computing hardware, are deployed at the individual device or instrument level. Distributed computing now pairs with distributed data. Through an internet of ML/AI-enhanced instruments, each ML/AI-enhanced instrument can be optimized for a specific purpose such as data reduction and real-time data processing. The large volumes, varieties, and generation rate of X-ray data
motivate automated processing and reduction in light sources, such as synchrotrons and XFELs, to reduce the memory requirement, minimize latency related to data transmission and processing, and lower energy and power consumption. 

Centralized approaches and large models are commonly executed by a large team of people. A Meta AI research team recently introduced the model called Segment Anything Model (SAM) and a dataset of  more than 1 billion masks on 11 Million images \cite{SAM:2023}. Nvidia unveiled Project Clara at its recent GTC conference, showing early results using DL post-processing to dramatically enhance existing, often grainy and indistinct echocardiograms (sonograms of the heart). Clara motivates acceleration in research 
being done on several fronts that exploits explosive growth in DL computational capability to perform analysis that was previously impossible or far too costly. One technique is called 3D volumetric segmentation that can accurately measure the size of organs, tumors or fluids, such as the volume of blood flowing through arteries and the heart. NVIDIA claims that a recent implementation, an algorithm called V-Net, “would’ve needed a computer that cost \$10 million and consumed 500 kW of power".


There are some other successes using ML and AI in areas such as HEP experiments (i.e. Higgs boson discovery) and electron microscopy. The discovery of the Higgs boson is a major challenge in HEP and can be setup as a classification problem. Many ML methods such as decision trees, logistic regression, and DL algorithms have been applied to solve the signal separation problem \cite{adam2015higgs, azhari2020higgs}. Meanwhile, ML and AI in electron microscopy are proposed to enable autonomous experimentation. Specifically, the automation of routine operations including but not limited to probe conditioning, guided exploration of large images, optimized spectroscopy measurements, and time-intensive and repetitive operations \cite{kalinin2021automated}. Edge ML and edge AI have already attracted a lot of attention in medicine. The fusion of DL and medical images creates dramatic improvements \cite{Marko:2018}. The concept is similar to techniques like high-dynamic range (HDR) photography, digital remastering of recordings or even film colorization in that one or more original sources of data are post-processed and enhanced to bring out additional detail, remove noise or improve aesthetics. 
 
We will give an overview of DL methods for real-time radiation image analysis as well as hardware solutions for DL acceleration at the edge. Specifically, this paper is organized as follows. In Section 2, we discuss different radiation detectors and imaging devices, the resulting big data generation at photon sources, and the motivations for edge computing and DL. In Section 3, we present an overview of popular neural network architectures and several image processing tasks that have potential to be performed on edge devices. In addition, we discuss examples of DL-based methods for each. In Section 4, we present on overview of hardware solutions for DL acceleration and recent works that have applied them for computing at the edge. Lastly, Section 5 concludes this paper.

\section{Experimental data generation at photon sources}

Data science at light sources is centered around scientific data generation and processing. Scientific data at synchrotrons and XFEL sources consist of experimental data, simulation and synthetic data, and meta data, such as detector calibration data, material properties of objects and sensors, and point spread functions of the detectors. Methods (imaging modalities) and detectors to collect experimental data are driven by the light sources, which continue to improve in source brightness, repetition rate, source coherence, photon energy, and spectral tunability. Computing hardware and algorithms are used to process experimental data and for data visualization. Computing hardware and algorithms are also used to simulate the experiments and produce synthetic data as close to the experimental data as possible for experimental data interpretation. Diversity of the materials to be integrated and imaged, together with the photon source and detector improvement have demanded continued improvements in computing hardware and algorithms towards real-time data processing, reductions in data transmission over long distances, and reducing data storage volumes.

\begin{figure}[t]
\centering
\includegraphics[width=0.6\linewidth]{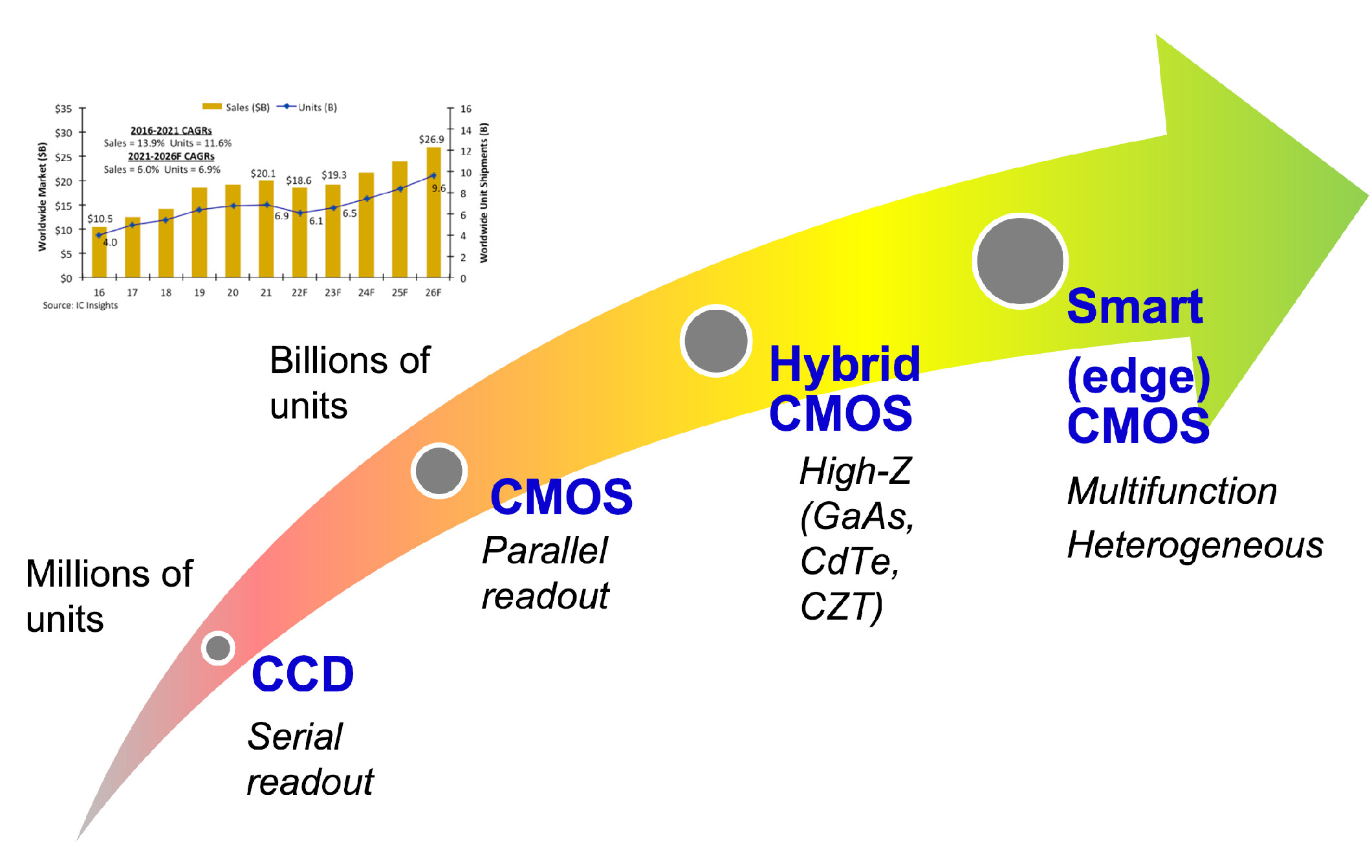}
\caption{Evolution of digital image sensor technology, which started with the introduction of the charge-coupled device (CCD) in the late 1960s. The latest trend is smart multi-functional CMOS image sensors enabled by three-dimensional (3D) integration in fabrication, innovations in heterogeneous materials and structures, neural networks, and edge computing. The upper left plot shows the increasing growth of sales for CMOS image sensors from 2016 to a projected value in 2026 \cite{ICinsights}.}
\label{fig:1}
\end{figure}

\subsection{Radiation Detectors and Imaging for Photon Science \label{sec:det}}

Complementary metal-oxide semiconductor (CMOS) pixelated detectors are now widely used in photon science, replacing charge-coupled devices (CCDs) as the primary digital imaging technology, see Figure \ref{fig:1}. Particle nature of photons motivates digitized detectors for photon counting. However, several factors complicate photon counting implementation in high-luminosity X-ray sources. The intensity of the sources can be too high to count individual photons one by one. The amount of X-ray photon-induced charge in CMOS detectors, which is the basis of X-ray photon counting, is not constant for the same X-ray energy. The source energy is not monochromatic, especially in imaging applications. Inelastic scattering of mono-chromatic X-rays can result in a broad distribution of X-ray photon energies after scattering by the object. When an optical camera is used together with a X-ray scintillator, see Table~\ref{tab1:data}, the energy resolution of individual X-rays based on the photon detection is worse than direct detection when the X-ray directly deposits its energy in a silicon photo-diode.

\subsection{Imaging Modalities }

X-ray microscopy uses X-ray lenses, zone plates, mirrors and other optics to modulate the X-ray field to form images \cite{niemann1976x}. As the X-ray intensities generated by synchrotrons and XFELs continue to increase, the advances in computational imaging modalities and lens-less X-ray modalities are increasingly used in synchrotrons and XFELs. In some cases, lens-less modalities may be preferred to avoid damages to X-ray lenses and mirrors. Lens-less modalities may also avoid aberration, diffraction due to imperfect X-ray lens, defects in zone plates, and other optics. The simplest lens-less X-ray imaging setup is radiography or projection imaging, pioneered by R\"ontgen. R\"ontgen’s lens-less radiographic imaging modality directly measures attenuated X-ray intensity due to absorption. Synchrotrons and XFELs also allow a growing number of phase contrast imaging, see Ref. \cite{wang2023ultrafast} and references therein. Other modalities include in-line holography \cite{spanne1999line} and coherent diffractive imaging \cite{miao2015beyond}. Additional phase and intensity modulation using pinholes, coded apertures, and kinoforms are also possible. Combinatorial X-ray modalities have also been introduced. For example, X-ray ptychography microscopy combines raster scanning X-ray microscopy with coherent diffraction imaging \cite{pfeiffer2018x}. Compton scattering, usually ignored in the synchrotron and XFEL setting, may offer some additional information about the samples and potentially reduce the dose required \cite{villanueva2018dose}. The versatility of modalities requires different off-line and real-time data processing techniques. Background reduction is a common issue for all X-ray modalities. Real-time data processing, including energy-resolving detection, is highly desirable to distinguish different sources of X-rays since the detector pixel may simultaneously collect X-ray photons from different sources of X-ray attenuation and scattering.

\subsection{Real Time In-pixel Data-processing} \label{sec:realtime_pixel}

When an X-ray photon is detected directly or indirectly through the use of a scintillator, charge-hole pairs are created through photo-to-electric conversion, or the photoelectric effect, within pixels of a camera or a pixelated array. CCD cameras, CMOS cameras, and low-gain avalanche detector (LGAD) arrays are now available for synchrotron and XFEL applications. Unlike a CCD camera, a CMOS image sensor collects charge and stores it in capacitors in pixels in parallel. Parallel charge collection and capacitor voltage digitization, which turns analog voltage signals into digitized signals, allow CMOS image sensors to operate at a much higher frame rate than CCDs. Charge and voltage amplification, in LGADs and sometimes in CMOS image sensors, are also used to improve signal-to-noise ratio. Any source of charge or voltage modulation not related to the photoelectric effect is a potential source of noise. The photoelectric effect itself can lead to so-called Poisson noise due to the probabilistic process of photo-to-electric conversion. Other sources of noise include thermal noise or dark current, salt’n’pepper noise (due to charge migration in and out of pixel defects and traps), and readout noise. 

\begin{figure}[t]
\centering
\includegraphics[width=0.6\linewidth]{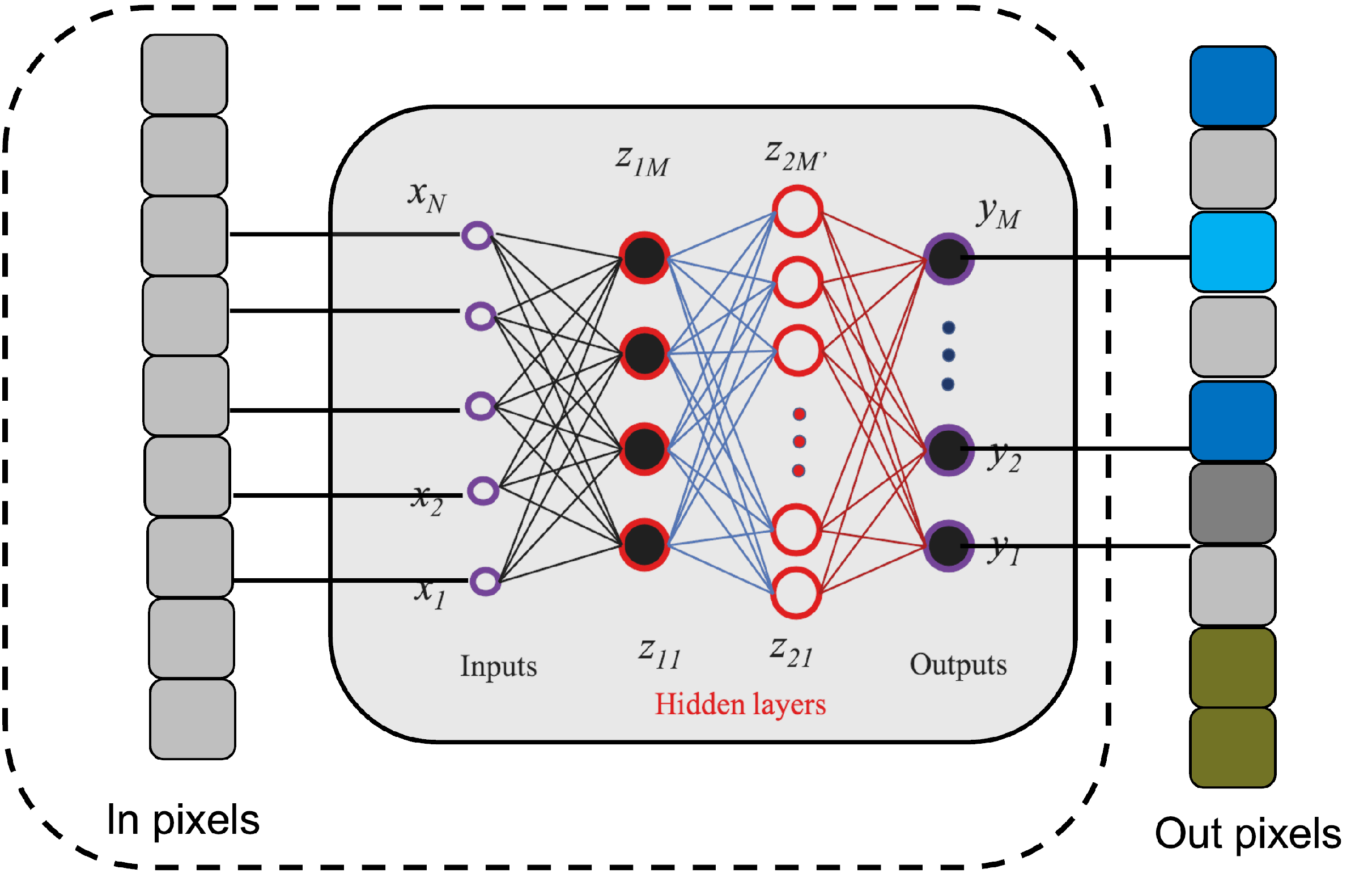}
\caption{Integrated neural network and image sensor for real-time data processing and reductions.}
\label{fig:phicam}
\end{figure}

Automated real-time in-pixel signal and data processing are therefore required in CMOS and other pixelated array sensors for noise rejection, noise reduction, and noise correction for charge and voltage amplification controls, and for charge sharing corrections. Figure \ref{fig:phicam} illustrates an example of an integrated neural network and image sensor for real-time data processing. If uncorrected, noise can corrupt the image information and make it hard for post processing or misleading for data interpretation. Charge and voltage amplification may lead to nonlinear distortion between the X-ray flux and voltage signal. When the X-ray flux is too high, the so-called plasma effect may also need correction. Charge-sharing happens when an X-ray photon arrives at a pixel border and the electron-hole pairs created are spread across multiple neighboring pixels. 

By using transistor circuits, correlated double sampling (CDS) is an extremely successful example in noise reduction. Adaptive gain control circuits have been implemented in the AGIPD high-speed camera \cite{allahgholi2019adaptive}. While real-time pixel-level signal processing by novel transistor circuits is important, there is also room for novel data-processing approaches that do not require hardware modifications to the pixels. As a recent example \cite{lin2023demonstration}, a physics-informed neural network was demonstrated to improve spatial resolution of neutron imaging. Other novel applications of neural networks and their integration with hardware, see Figure \ref{fig:phicam}, may offer new possibilities in noise reduction and image corrections. Integrated hardware and software (neutral networks are emphasized here) approaches for optimal performance also need to take into account of the complexity of the workflow \cite{Orp:1994,MM:2018,FSN:2022}, or computational cost, power consumptions, constrained by the frame rate and other metrics. For example, the computational cost of an $n\times n$ matrix is $O(n^3)$ \cite{GL:2013}. 

%

\begin{table*}[!htbp]
\small
\centering
\caption{\label{tab1:data}%
\small A comparison of different camera data rates, additional details and examples may be found in \cite{wang2023ultrafast}. The state of the art is  $>$10 Gpixel/s in continuous mode imaging. The burst mode imaging is $>$ 1 Tpixel/s \cite{turchetta2017towards}.}
\begin{threeparttable}
\begin{tabularx}{\linewidth}{@{\extracolsep{\fill}}lccccc} \hline
\textrm{Detector} &
\textrm{Facility} &
\textrm{Array format}&
\textrm{frame-rate } &
\textrm{data} &
\textrm{Data rate}
\\
 (camera) & (particle/ &
\textrm{(voxel size, $\mu$m$^3$ /}&
\textrm{(fps/} & bits & (GB/s)
\\
 &  photon)&
\textrm{ pixel size, $\mu$m$^2$ )}&
\textrm{Hz)} & &
\\ \hline

AGIPD~\cite{allahgholi2019megapixels} & Eu-XFEL &  512 $\times$ 128~\tnote{a} & 16 k/6.5 M~\tnote{b} & 14 & 1.85\\
& (12.4 keV)&(200$^2 \times$ 500 )&& &\\
CS-PAD~\cite{philipp2011pixel} & LCLS &  194 $\times$ 370~\tnote{c} & 120 & 14 & 0.02\\
&(8.3 keV)&(110$^2 \times$ 500 )&& &\\
ePix100~\cite{carini2016epix100}  & LCLS &  384 $\times$ 352~\tnote{d} &  120 &  14 & 0.03\\
&(8.3 keV)&(50$^2 \times$ 500)& ($\leq$ 240) & &\\
ePix10k  & LCLS &  384 $\times$ 352~\tnote{e} &  120 &  14 & 0.03 \\
&(8.3 keV)&(100$^2 \times$ 500)& ($\leq$ 10$^3$) & &\\
EIGER2  & APS \& others & 1028 $\times$ 512~\tnote{f} &  2.25 k &  16 & 2.37\\
(Dectris) & & (75$^2 \times$ 450)& (4.5 k) & (8) & \\
HEXITEC~\cite{veale2018hexitec} & DIAMOND& 80$^2$& 6.3 - 8.9 k & 14 & 0.07 - 0.1\\
&(2-200 keV) &(250$^2 \times$ 1000~\tnote{g})&& & \\
Icarus~\cite{claus2017design} & NIF, Z & 1024 $\times$ 512 & $\geq$ 250 M~\tnote{h} & 10 & 163,840\\
(Advanced & (0.7 - 10 keV) &(25$^2 \times$ 25)& & & \\
hCMOS Sys.) &  && & & \\
JUNGFRAU~\cite{leonarski2020jungfrau} & PSI & 1024 $\times$ 512 & 2.2k & 16 & 2.31\\
& (12 keV)  && & & \\
MM-PAD~\cite{gadkari2022characterization} & CHESS &  128$^2$& 10 k/100 M~\tnote{i} & 14 & 0.3 - 2867 \\
&  ($>$20 keV)~\tnote{j} & (150$^2 \times$ 500) & & &\\
SOPHIAS & SACLA &  891 $\times$ 2157 & 60 & 12 & 0.17 \\
& & (30$^2 \times$ 500)& & & \\ \hline
HPV-X2~\cite{tochigi2012global} & APS \& others &  400 $\times$ 250 & 7.8 k/5 M~\tnote{k} & 10 & 0.98 - 625 \\
(Shimadzu) &  (10-40 keV) &  (32$^2$)& & & \\
Kraken~\cite{lewis2021new} & NNSS & 800 $\times$ 800 & 20 M~\tnote{l} & 12 & 19,200 \\
 &  & (30$^2$) & & & \\
MX170-HS & LCLS  & 3840$^2$ & 2.5~\tnote{m} & 16 & 0.07 \\
(Rayonix) &  (8-12 keV)  & (44$^2$)& & & \\
PI MAX 4 & APS  &  1024$^2$ & 26~\tnote{n} & 16 & 0.05 \\
(Teledyne) &  (10-40 keV) &  (12.8$^2$)& & & \\
 \hline
\end{tabularx}
\begin{tablenotes}
	\item[a] AGIPD is deployed as mega-pixel/voxel cameras through tiling.
	\item[b] Burst mode for 352 stored frames.
	\item[c] CS-PAD is deployed as tiled 2, 8, and 32 modules with up to 2.3 M voxels.
	\item[d] ePix100 is deployed as tiled 4 modules with about 0.5 M voxels. 
	\item[e] ePix10K replaces CS-PAD, and  is deployed as a single, or tiled 16 modules with about 2.2 M voxels. 
	\item[f] Eiger2 is deployed as a single, or tiled modules with more than 10 M voxels.
	\item[g] Also 2 mm CdZnTe
	\item[h] In burst mode for 4 frames.
	\item[i] In burst mode for 8 frames.
	\item[j] When using 750 $\mu$m CdTe as sensor
	\item[k] In burst mode for 128 stored frames; or 10 M Hz frame rate and 256 stored frames possible by reducing the number of pixels by half.
	\item[l] In burst mode for 8 frames. Read noise 157 $e^-$, Full Well 4.0$\times$10$^5$ $e^-$. Buttable to larger array 2$\times$2.
	\item[m] Higher frame rate can be obtained through pixel binning, at 10$\times$10 binning, the frame rate increases to 120 Hz
	\item[n] Higher frame rate can be obtained through pixel binning, at 4$\times$4 binning, the frame rate increases to 95 Hz
\end{tablenotes}
\end{threeparttable}
\end{table*}

\section{Deep Learning for Image Processing}

\begin{figure}[!tbh]
\centering
\includegraphics[width=0.9\linewidth]{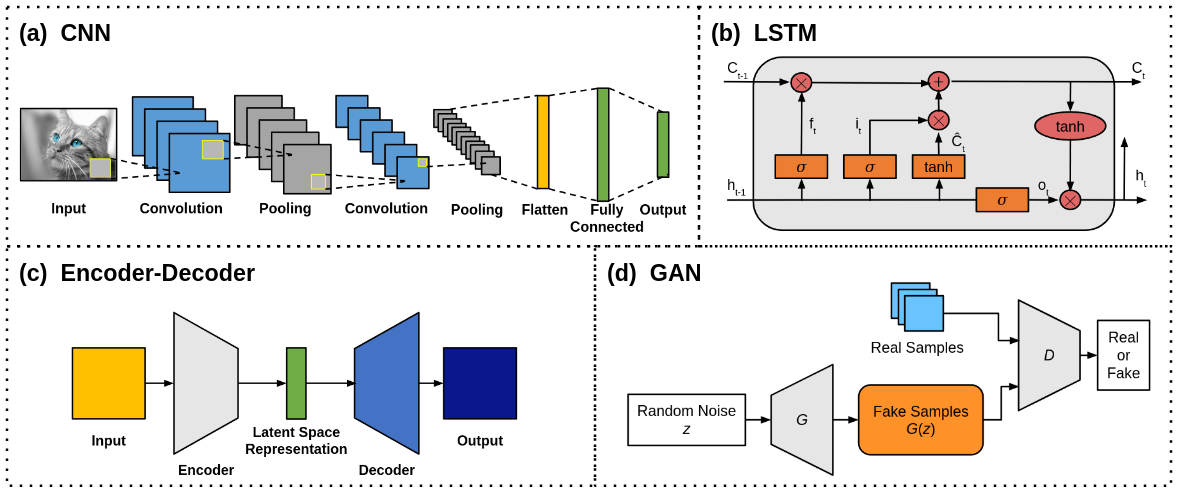}
\caption{Basic neural network architectures for (a) CNNs, (b) LSTMs, (c) encoder-decoders, and (d) GANs.}
\label{fig:nn}
\end{figure}

In recent years, deep learning (DL) has contributed significantly to the progress in computer vision, especially in different areas of image processing tasks including but not limited to image denoising, segmentation, super-resolution, and classification. DL is a sub-field of ML and AI that utilize neural networks (NNs) and their superior nonlinear approximation capabilities to learn underlying structures and patterns within high-dimensional data \cite{lecun2015deep}. In other words, DL aims to learn multiple levels of representation, corresponding to a hierarchy of features or concepts, where higher-level features are defined from lower-level ones and lower-level ones can help build up higher-level features. 

\subsection{Neural Network Architectures}

This section provides an overview of different popular deep neural network (DNN) architectures used for image processing tasks. These widely used architectures include but are not limited to convolutional neural networks (CNNs), long short term memory (LSTM), encoder-decoder networks, and generative adversarial networks (GANs). Due to space limitations, other DNN architectures such as transformers \cite{khan2022transformers}, restricted boltzamann machines \cite{zhang2018overview}, and extreme learning \cite{cao2015extreme} will not be covered here.

\subsubsection{Convolutional Neural Networks (CNNs)}

CNNs are one of the most widely used architectures in DL, especially for image processing tasks, due to its inherenet spatial invariance property. The built-in convolutional layers allow the network to naturally reduce the high dimensionality of the input data, i.e. images, without information loss. Figure \ref{fig:nn}(a) shows the basic architecture of CNNs, which usually consists of 3 types of layers: i) convolutional layers, ii) pooling layers, and iii) fully connected layers. The convolutional layer uses various kernels to convolve the entire input image, including intermediate feature maps, and generate new feature maps. There are 3 major advantages of the convolutional operation \cite{zeiler2013hierarchical}: i) the number of parameters is reduced by using weight sharing mechanisms, ii) the correlation among neighboring pixels are easily learned through local connectivity, and iii) the location of objects are fixed due to spatial invariance. Generally following a convolutional layer, the pooling layer is used to further reduce the dimensions of feature maps and network parameters. The average pooling and max pooling methods are commonly used, and their theoretical performances have been evaluated by \cite{boureau2010theoretical} and \cite{scherer2010evaluation}, where max pooling is shown to achieve faster convergence and improved CNN performance. Lastly, the fully connected layers follows the last pooling or convolutional layer to convert the 2D feature maps into a 1D vector for additional feature mapping, i.e. labels. A few of the well known CNN models are the AlexNet \cite{krizhevsky2012imagenet}, VGG \cite{simonyan2014very}, GoogLeNet \cite{szegedy2015going}, and ResNet \cite{he2016deep}, where all the models were top 3 finishers in the ImageNet Large Scale Visual Recognition Challenge (ILSVRC).

\subsubsection{Long-Short Term Memory (LSTM)}

LSTMs \cite{yu2019review} are a special type of recurrent neural network (RNN) that is commonly used to process sequential datasets, such as audio recordings, videos, and time-series data. Figure \ref{fig:nn}(b) shows the basic structure of a LSTM block, which consists of 3 gates (input, output, and forget gates) that regulate the stored memory and information flow within the block. The multiple gate architecture of LSTMs is specifically designed to capture long-term dependencies in the data as well as to avoid the vanishing gradient problem of vanilla RNNs \cite{yu2019review}. Other LSTM architectures are derived from the basic architecture in Figure \ref{fig:nn}(b) such as LSTM without a forget gate, LSTM with peephole connections, the gated recurrent unit (GRU), and other variants \cite{yu2019review}.

\subsubsection{Encoder-Decoders}

Encoder-decoder neural networks, also known as sequence-to-sequence networks, are a type of network that learns to map the input domain to a desired output domain \cite{Goodfellow-et-al-2016}. As shown in Figure \ref{fig:nn}(c), the network consists of two main components: an encoder network which uses an encoder function $h=f(x)$ to compress the input $x$ into a latent-space representation $h$, and a decoder network $y=g(h)$ that produces a reconstruction $y$ from $h$. The latent-space representation $h$ prioritizes learning the important aspects of the input $x$ which are useful in reconstructing the output $y$. A special case of encoder-decoder models, autoencoders are networks in which the input and output domains are the same. These networks are popularly used in DL applications involving sequence-to-sequence modeling such as natural language processing \cite{bahdanau2014neural}, image captioning \cite{herdade2019image}, and speech recognition \cite{chiu2018state}.

\subsubsection{Generative Adversarial Networks (GANs)}

GANs \cite{goodfellow2014generative} are increasingly popular DL frameworks for generative AI models. Classical GANs consist of 2 different networks, a generator and a discriminator, as shown in Figure \ref{fig:nn}(d). The generator network $G$ aims to generate data that is indistinguishable from the real data by learning a mapping from an input noise distribution $z$ to a target distribution $y$. Meanwhile, the discriminator network $D$ takes as input the real and generated data, and aims to correctly classify them as ``real'' or ``fake'' (generated). The GAN learning objective takes on a game-theoretic approach as a two player minimax game between $G$ and $D$. Let $\mathcal L$ denote the loss function and the GAN objective as $\min_G ~ \max_D ~ \mathcal L (G,D)$. Intuitively, $D$ aims to minimize its own classification error, which maximizes $\mathcal L (G,D)$. Meanwhile, $G$ aims to maximize the classification error of $D$, which minimizes $\mathcal L (G,D)$. This adversarial loss function allows both models to be trained simultaneously and in competition with each other. Other GAN architectures are derived from the basic architecture in Figure \ref{fig:nn}(d) such as conditional GANs, GANs with inference models, and adversarial autoencoders \cite{creswell2018generative}.

\subsection{Image Processing Techniques}

This section provides an overview of several image processing tasks that have potential to be performed on edge devices. In addition, this section surveys different works that have applied  the DL-based image processing techniques to radiographic image processing.

\subsubsection{Restoration}

Image restoration is the process of adjusting the quality of digital images such that the enhanced image can facilitate further image analysis. Common enhancement operations include histogram-based equalization, brightness, and contrast adjustment. However, these operations are very elemental and advanced operations are necessary to further improve the perceptual quality. These advanced operations include image denoising, deblurring, and super-resolution (SR). 

\textbf{Denoising.} One of the fundamental challenges in image processing, image denoising aims to estimate the ground-truth image by suppressing internal and external noise factors such as sensor and environmental noise, as discussed in Section \ref{sec:realtime_pixel}. Conventional methods including but not limited to adaptive nonlinear filters, Markov random field (MRF), and weighed nuclear norm minimization (WNNM), have achieved good performance in image denoising \cite{tian2020deep}, however, they suffer from several drawbacks \cite{lucas2018using}. Two major drawbacks are the need to manually set parameters as the proposed methods are non-convex and the high computational cost for the optimization problem for the test phase. To overcome these challenges, DL methods are applied for image denoising problems to learn the underlying noise distribution. Various neural network architectures, such as CNNs, encoder-decoders, and GANs, have been proposed for image denoising in recent years; see \cite{tian2020deep} for details.

On example application that uses image denoising is in X-ray computed tomography (CT). X-ray CT imaging is a common noninvasive imaging technique that allows for reconstructing the internal structure of objects by using 3D reconstruction from 2D projection images; see Section \ref{sec:3d_reconstruction} on 3D reconstruction. The spatial resolution of CT images can range from tens of microns to a few nanometers, while higher resolutions can be obtained by using higher radiation doses. However, some experiments may require short exposure times or low radiation dosage to avoid damaging the sample. The low-dose image conditions results in noisy 2D projection images, which in turn impacts the quality of the 3D reconstructed image. To address this issue, \cite{liu2020tomogan} developed a GAN-based image denoising method called TomoGAN. TomoGAN is a conditional GAN model where the generator $G$ conditionally uses the noisy reconstruction as input and outputs enhanced (denoised) reconstructions. Furthermore, the generator network architecture adopts a modified U-Net \cite{ronneberger2015u} architecture, popularly used for image segmentation. Meanwhile, the discriminator $D$ is trained to classify reconstructions of the enhanced reconstructions and reconstructions of normal dose projections. \cite{liu2020tomogan} evaluates the effectiveness of TomoGAN on two experimental (shale sample) datasets. TomoGAN outperforms conventional methods in noise reduction and reports a higher structural similarity (SSIM) value. In addition, TomoGAN is demonstrated to be robust to images with dynamic features from faster experiments, e.g. collecting fewer projections and/or using shorter exposure times.

Denoising has also been applied to synchrotron radiation CT (SR-CT) in a recent work by \cite{duan2023sparse2noise}, which developed a CNN-based image denoising method called Sparse2Noise. Similar to the previous work for TomoGAN, this work presents a low-dose imaging strategy and utilizes paired normal-flux CT images (sparse-view) and low-flux CT image (full-view) to train Sparse2Noise. In addition, Sparse2Noise also adopts a modified U-Net architecture for its performance of removing image degradation factors such as noise and ring artifacts. The Sparse2Noise network takes as input the normal-flux CT images into the modified U-Net architecture and outputs the enhanced image. During training, the network is trained in a supervised fashion using the low-flux CT images. The loss function to update the network weights is defined to minimize the difference between the enhanced image and the reconstructed low-flux CT image. \cite{duan2023sparse2noise} evaluates the effectiveness of Sparse2Noise on one simulated and two experimental datasets. Furthermore, Sparse2Noise is compared to simultaneous iterative reconstruction technique (SIRT), unsupervised deep imaging prior (DIP), and supervised training algorithms Noise2Inverse \cite{hendriksen2021deep} and Noise2Noise \cite{lehtinen2018noise2noise}. For the simulated dataset, Sparse2Noise outperforms all methods by achieving the highest SSIM and peak signal to noise ratio (PSNR) values, and in terms of removing image degradation factors such as noise and ring artifacts. For the experimental datasets, Sparse2Noise also achieves the best performance in terms of noise and ring artifact removal. Most importantly, however, Sparse2Noise can achieve excellent performance for low-dose experiments (0.5 Gy per scan).

\textbf{Deblurring.} Image deblurring aims to recover a sharp image from a blurred image by suppressing blur factors such as lack of focus, camera shake, and target motion. Some blur factors are application specific such as multiple Coulomb scattering and chromatic aberration in proton radiography \cite{morris2013charged}. A blurred image can be modeled mathematically as $B = K*I +N$, where $B$ denotes the blurred image, $K$ the blur kernel, $I$ the sharp image, $N$ the additive noise, and $*$ the convolution operation. The blur kernel $K$ is typically modeled as a convolution of blur kernels that are spatially invariant or varying \cite{zhang2022deep}. Conventional methods aim to solve the inverse filtering problem to estimate $K$, however, this is an ill-posed problem as the sharp image $I$ needs to be estimated as well. To address this issue, prior-based optimization approaches, also known as maximum a posteriori (MAP)-based approaches, have been proposed to define priors for $K$ and $I$ \cite{biyouki2023comprehensive}. While these approaches are shown to achieve good results for image deblurring, deep learning approaches can further improve the accuracy of the blur kernel estimation or even skip the kernel estimation process altogether by using end-to-end methods. Various neural network architectures, such as CNNs, LSTMs, and GANs, have been proposed for image deblurring; see \cite{zhang2022deep,biyouki2023comprehensive} for details.

One example application that uses image deblurring is in neutron imaging restoration (NIR), a non-destructive imaging method. However, the neutron images suffer from noise and blur artifacts due to the neutron source and the digital image system. The low quality of raw neutron images limits their applications in research, and thus image denoising and deblurring techniques are necessary to produce sharp images. To address these issues, \cite{yang2023deep} proposes a fast and lightweight neural network called DAUNet. DAUNet consists of three main blocks: a feature extraction block (FEB), multiple cascaded attention U-Net blocks (AUB), and a reconstruction block (RB). First, DAUNet takes as input a degraded neutron image and feeds it into the FEB to extract important underlying features. Next, the AUB inputs the extracted feature maps into a modified U-Net with an attention mechanism, which allows U-Net to focus on harder to address features such as texture and structure information, and outputs a restored image. Last, the RB block outputs the enhanced image by reconstructing the restored image. To evaluate DAUNet, its performance is compared with several popular DNN image restoration methods such as DnCNN \cite{zhang2017beyond} and RDUNet \cite{gurrola2021residual}. Due to the lack of available neutron imaging datasets, the networks are trained on X-ray images that are similar to the neutron imaging principle; specifically, the X-ray images are obtained from the SIXray dataset \cite{miao2019sixray}, where 4699 and 23 images are used as the training and test set respectively. In addition, seven clean neutron images are added to the test set. Results show that DAUNet can effectively improve the image quality by removing noise and blurring artifacts, while achieving quality close to the large network with faster running times and a smaller number of network parameters.

\textbf{Super-resolution (SR).} Image SR is the process of reconstructing high-resolution images from low-resolution images. It has been widely applied in many real-world applications, especially in medical imaging \cite{li2021review} and surveillance \cite{jiang2021deep}, where the spatial resolution of captured images are not sufficient due to limitations such as hardware and imaging conditions. A variety of DL-based methods for SR have been explored, ranging from CNN-based methods (e.g SRCNN \cite{dong2015image}) to more recent GAN-based methods (e.g. SRGAN \cite{ledig2017photo}). In addition to utilizing different neural network architectures, DL-based SR algorithms also differ in other major aspects such as their loss functions and training approaches \cite{wang2020deep}. These differences result from various factors that contribute to the degradation of image quality including but not limited to blurring, sensor noise, and compression artifacts. Intuitively, one can think of the low-resolution image as the output of a degradation function with an input high-quality image. In the most general case, the degradation function is unknown and an approximate mapping is learned through deep learning. These degradation factors influence the design of loss function, and thus training approaches. A detailed discussion of the various loss functions, SR network architectures, and learning frameworks is out of scope for this paper; however, see \cite{wang2020deep} for details.

An example application that applies super resolution is for X-ray CT imaging. As mentioned earlier, CT imaging has many factors that impact the resulting image quality such as radiation dose and slice thickness. In addition, 3D image reconstruction may require heavy computational power due to the number of slices or projection views taken, where thicker slices results in lower image resolution, and slower operational speed, which increases with the number of slices. To address this issue, it is desirable to obtain higher-resolution (thin-slice) images from low-resolution (thick-slice) ones. \cite{park2018computed} develops an end-to-end super-resolution method based on a modified U-Net. The network takes as input the low-resolution image and outputs the high-resolution one. The network is trained on slices of brain CT images obtained from a 65 clinical positron emission tomography (PET)/CT studies for Parkinson's disease. The low-resolution images are generated as the moving average of five high resolution slices and the ground-truth image is taken as the middle slice. The performance of the proposed method is compared with the Richardson-Lucy (RL) deblurring algorithm using the PSNR and normalized root mean square error (NRMSE) metrics. The results show that the proposed method achieves the highest PSNR and lowest NRMSE values compared to the RL algorithm. In addition, the noise level of the enhanced images are reported to be lower than that of the ground-truth.

\subsubsection{Segmentation}

Image segmentation is the process which segments an image or video frames into multiple regions or clusters, where each pixel can be represented by a mask or be assigned a class \cite{szeliski2022computer}. This task is essential in a broad range of computer vision problems, especially for visual understanding systems. A few applications that utilize image segmentation include but are not limited to medical imaging for organ and tumor localization \cite{wang2022medical}, autonomous vehicles for surface and object detection, and video footage for object and people detection and tracking \cite{badrinarayanan2017segnet}. Numerous techniques for image segmentation have been proposed throughout the years, ranging from early techniques based on region splitting or merging such as thresholding and clustering algorithms, to newer algorithms based on active contours and level sets such as graph cuts and Markov random fields \cite{szeliski2022computer,minaee2021image}. Although these conventional methods have achieved acceptable performance for some applications, image segmentation still remains a challenging task due to various image degradation factors such as noise, blur, and contrast. To address these issue, numerous deep learning methods have been developed and have been shown to achieve remarkable performance. This is due to the powerful feature learning capabilities of DNNs, which allows DNNs to have reduced sensitivity to image degradation factors compared to the conventional methods. Popular neural network architectures used for DL-based segmentation includes CNNs, encoder-decoder models, and multiscale architectures; see \cite{minaee2021image} for details. Two popular DNN architectures used for image segmentation problems are U-Net \cite{ronneberger2015u} and SegNet \cite{badrinarayanan2017segnet}.

Image segmentation is an important step in analyzing X-ray radiographs from, for example, inertial confinement fusion (ICF) experiments 
\cite{falato2022contour}. ICF experiments typically use single or double shell targets which are imploded as the laser energy or laser-induced X-rays rapidly compress the target surface. X-ray and neutron radiographs of the target provide insight to the shape of target shells during the implosion. Contour extraction methods are used to extract the shell shape to conduct shot diagnostics such as quantifying the implosion and kinetic energy, identifying shell shape asymmetries, and determining instability information. \cite{falato2022contour} uses U-Net \cite{ronneberger2015u}, a CNN architecture for image segmentation, to output a binary masked image of the outer shell in ICF images. The shell contour is then extracted from the masked image using edge detection and shape extraction methods. Due to the limited number of actual ICF images, a synthetic dataset consisting of 2000 experimental-like radiographs is used to train the U-Net. In addition, the synthetic dataset provides ground-truth ICF image-mask pairs, which are required to train U-Net. The trained U-Net is tested on experimental images and has successfully extracted the binary mask of high-signal-to-noise ratio ICF images.

Another example of X-ray image segmentation is for 
the Magnetized Linear Inertial Fusion (MagLIF) 
experiments at Sandia National Laboratory's Z-facility \cite{lewis2022statistical}. The MagLIF experiments compresses a cylindrical beryllium tube, also known as a liner, filled with pure deuterium fuel using a very large electric current on the order of $O(20MA)$. Before compression, the deuterium fuel is pre-heated and an axially oriented magnetic field is applied. The electric current causes the liner to implode and compresses the deuterium fuel in a quasi-adiabatic implosion. The magnetic field flux is also compressed which aids in the trapping of charged fusion particles at stagnation. X-ray radiographs are taken during the implosion process for diagnostics and to analyze the resulting plasma conditions and liner shape. To better analyze the implosion, a CNN model is proposed to segment the captured X-ray images into fuel strand and background. The CNN is trained using synthetically generated and augmented dataset of 10,000 X-ray images and their corresponding binary masks. The trained CNN is tested on experimental images where the results generally demonstrate excellent fuel-background segmentation performance. The worst segmentation performance is due to factors such as excessive background noise and X-ray image plate damage.

\subsubsection{Compression} 

Image compression is the process of reducing the file size, in bytes, without reducing the quality of the image below a threshold. This process is important in order to save memory storage space and to reduce the memory bandwidth to transmit data, especially for running image processing algorithms on edge devices. The fundamental principle of compression is to reduce spatial and visual redundancies in images, by exploiting inter-pixel, psycho-visual, and coding redundancies. Conventional methods commonly leverage various quantization and entropy coding techniques \cite{mishra2022deep}. Popularly used conventional methods for lossy and lossless compression includes but are not limited to JPEG \cite{wallace1992jpeg}, JPEG2000, wavelet, and PNG. While conventional methods are widely used for both image and video compression, their performance is not the most optimal for all types of image and video applications. DL approaches can achieve improved compression results due to several factors. DNNs can learn non-linear mappings to capture the compression process as well as extract the important underlying features of the image through dimensionality reduction. For example, an encoder network or CNN can extract important features into a latent feature space for compact representation. In addition, DNNs can implement direct end-to-end methods using networks such as encoder-decoders to directly obtain the compressed image from an input sharp image. Furthermore, once a DNN is trained, the inference time is much faster. For DL-based image compression methods, the most commonly used neural network architectures are CNNs, encoder-decoders, and GANs \cite{mishra2022deep}.

\subsubsection{Sparse Sampling}

\begin{figure}[!bt]
\centering
\includegraphics[width=0.95\linewidth]{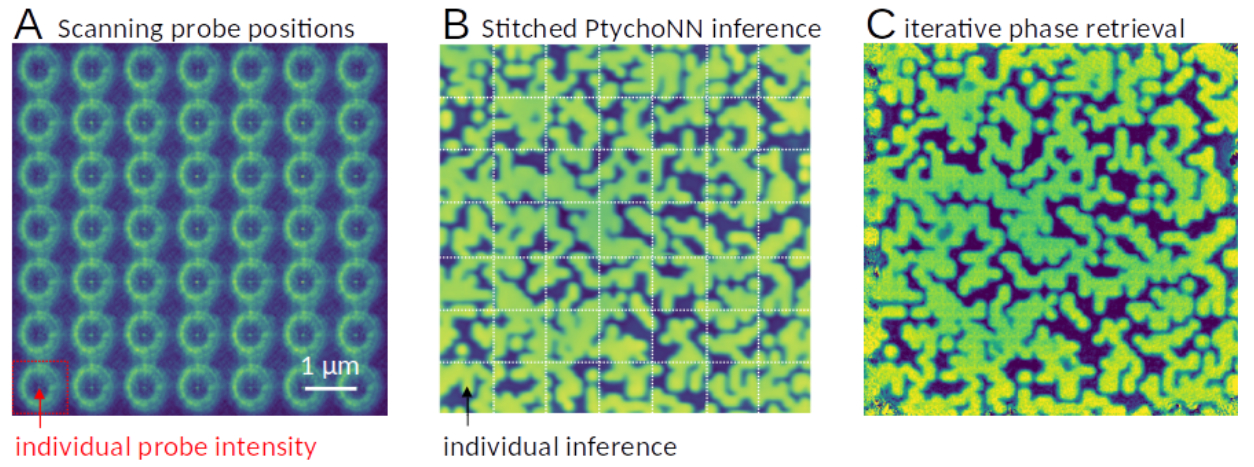}
\caption{Sparse-sampled single-shot ptychography reconstruction using PtychoNN. (a) scanning probe positions with minimal overlap. (b) Single-shot PtychoNN predictions on 25$\times$ sub-sampled data compared to (c) ePIE reconstruction of the full resolution dataset. }
\label{fig:ptychoNN}
\end{figure}

A closely related process to image compression is sparse sampling. While compression aims to reduce the file size, sparse sampling, also known as compressed sensing (CS), aims to efficiently acquire and reconstruct a signal by solving underdetermined linear systems. It has been shown in CS theory that a signal can be recovered from sampling fewer measurements than required by the Niquist-Shannon sampling theorem \cite{donoho2006compressed}. As a result, both memory storage space and data transmission bandwidth can be reduced. In conventional methods, CS algorithms need to overcome two main challenges: the design of the sampling and reconstruction matrices. Numerous methods have been proposed including but no limited to random and binary sampling matrices and reconstruction methods using convex-optimization and greedy algorithms \cite{shi2019image}. However, these conventional methods suffer from long computational times or low quality reconstruction. DL approaches allow for fast inference (reconstruction) times for a trained network, as well as learning non-linear functions for higher quality signal reconstruction \cite{shi2019image,machidon2023deep}.

Neural network (NN) models that learn to invert X-ray data have also been shown to significantly reduce the sampling requirements faced by traditional iterative approaches. For example, in ptychography, traditional iterative phase retrieval methods require at least 50\% overlap between adjacent scan positions to successfully reconstruct sample images as required by Nyquist-Shannon sampling. In contrast, Figure \ref{fig:ptychoNN}(b) shows image reconstructions obtained from PtychoNN when sampled at 25$\times$ less than required for conventional phase retrieval methods \cite{cherukara2020ai}. Figure \ref{fig:ptychoNN}(a) shows the probe positions and intensities, there is minimal overlap between probes. Through use of inductive bias provided through online training of the network \cite{babu2022deep}, PtychoNN is able to reproduce most of the features seen in the sample even when provided extremely sparse data. Figure \ref{fig:ptychoNN}(c) shows the same region reconstructed using an oversampled dataset and traditional iterative phase retrieval. 
Furthermore, \cite{babu2022deep} demonstrated live inference performance during a real experiment using an edge device and running the detector at its maximum frame rate of 2 kHz.

\begin{figure}[!tb]
\centering
\includegraphics[width=0.6\linewidth]{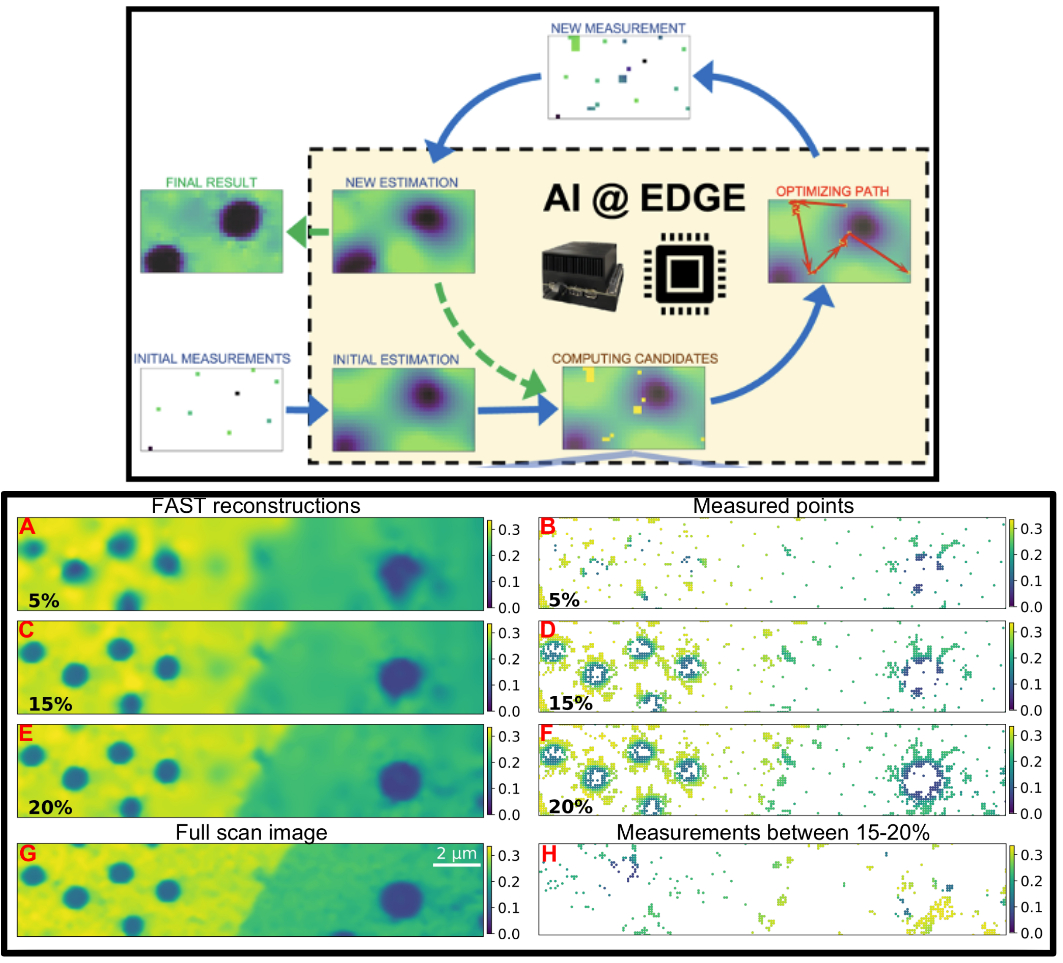}
\caption{FAST framework for autonomous experimentation. Top panel shows the workflow that enables real-time steering of scanning microscopy experiments. Bottom panel shows reconstructed images at 5\%, 15\% and 20\% sampling along with the corresponding locations from which they were sampled. In addition, the full-grid pointwise scan and corresponding points sampled between 15-20\% is also shown. Reproduced with permission from \cite{kandel2023demonstration}.}
\label{fig:fast}
\end{figure}

In the previous example, DL is used to reduce sampling requirements but not to alter the sampling strategy. In other words, the scan proceeds using a conventional acquisition strategy, but using fewer points along that trajectory than traditionally required. In contrast, active learning approaches are being developed that use data-driven priors to direct the acquisition strategy. Typically, this is treated as a Bayesian optimization (BO) problem using Gaussian processes (GPs). This method has been applied to a variety of characterization modalities including scanning probe microscopy \cite{vasudevan2021autonomous}, X-ray scattering \cite{noack2019kriging}, and neutron characterization \cite{venkatakrishnan2023adaptive}. A downside to such approaches is that the computational complexity typically increases as $O(N^3)$ with the action space \cite{liu2020gaussian}, making real-time decision a challenge. To address these scaling limitations which are critical especially in fast scanning instruments, recent work has demonstrated the use of pre-trained NNs to make such control decisions \cite{schloz2023deep,kandel2023demonstration}. Figure \ref{fig:fast} shows the workflow and results from the Fast Autonomous Scanning Toolkit applied to a 
scanning diffraction X-ray microscopy measurement of a WSe$_2$ sample. Starting from some quasi-random initial measurements, FAST generates an estimate of the sample morphology, predicts the next batch of 50 points to sample from, triggers acquisition on the instrument, analyzes the image after the next set of points has been acquired and continues the process until the improvement in sample image is minimal. Bottom panel of Figure \ref{fig:fast}, A,C,and E show the predicted image after 5\%, 10\% and 15\% sampling while Figure \ref{fig:fast} B, D, and F shows the points preferentially selected by the AI. The AI has learned to prioritize acquisition where the expected information gain is maximum, e.g. around contrast features on the sample.

\subsubsection{3D Reconstruction} \label{sec:3d_reconstruction}

Image-based 3D reconstruction is the process of inferring a 3D structure from a single or multiple 2D images, and is a common topic in the fields of computer vision, medical imaging, and virtual reality.  This problem is well known to be an ill-posed inverse problem. Conventional methods attempt to formulate a mathematical formula for the 3D to 2D projection process, use prior models, 2D annotations, and other techniques \cite{han2019image,fu2021single}. In addition, high quality reconstruction typically requires 2D projections from multiple views or angles, which may be difficult to calibrate (i.e. cameras) or time consuming to obtain (i.e. CT) depending on the application. DL techniques and the increasing availability of large datasets motivates new advances in 3D reconstruction by address challenges found in conventional methods. The popular networks used for image-based 3D reconstruction are CNNs, encoder-decoder, and GAN models \cite{han2019image}.

\begin{figure}
\centering
\includegraphics[width=0.8\linewidth]{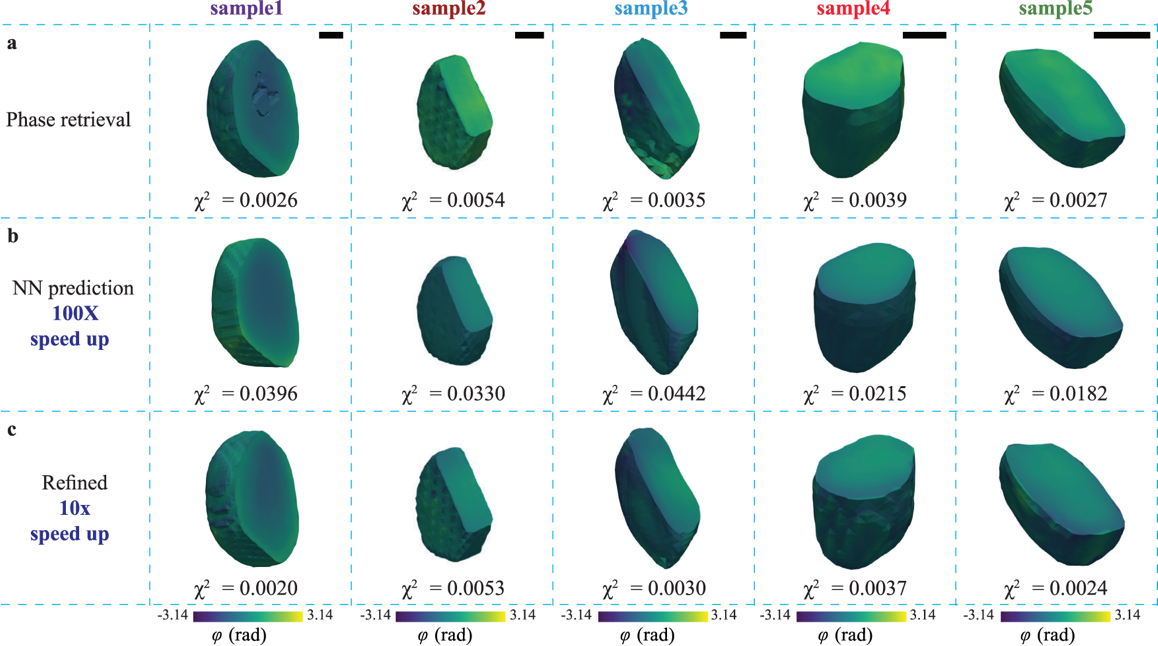}
\caption{Comparison of 3D sample images obtained by (a) phase retrieval, (b) AutoPhaseNN, and (c) AutoPhaseNN + phase retrieval. Reproduced with permission from \cite{yao2022autophasenn}.}
\label{fig:3D_comparison}
\end{figure}

X-ray phase information is now available for 3D reconstruction in the state-of-the-art X-ray sources such as synchrotrons and XFELs. In contrast to iterative phase retrieval methods that incorporate NNs through a DIP or other means, single-shot phase retrieval NNs provide sample images from a single pass through a trained NN. The inference time on a trained NN is minimal and such methods are hundreds of times faster than conventional phase retrieval \cite{guan2019ptychonet,cherukara2018real}. Figure \ref{fig:3D_comparison}(b) and \ref{fig:3D_comparison}(a) compare AutoPhaseNN and traditional phase retrieval for 3D coherent image reconstruction, respectively \cite{yao2022autophasenn}. AutoPhaseNN is trained to invert 3D coherently scattered data into sample image in a single shot. Once trained AutoPhaseNN is $>$100$\times$ faster that iterative phase retrieval with some reduction in accuracy. The prediction from AutoPhaseNN can also be used to seed phase retrieval, i.e. provide an initial estimate which can be refined by a few iterations of phase retrieval. This combined approach of NN + phase retrieval is shown to be both faster and more accurate than iterative phase retrieval. 

A recent work by \cite{scheinker2020adaptive} developed an adaptive CNN-based 3D reconstruction method for coherent diffraction imaging (CDI), a non-destructive X-ray imaging technique that provides 3D measurements of electron density with nanometer resolution. The CDI detectors record only the intensity of the complex diffraction pattern of the incident object. However, all phase information is lost in this detection method, and thus results in an ill-posed inverse Fourier transform problem to obtain the 3D electron density. Conventional methods encounter many challenges including expert knowledge, sensitivity to small variations, and heavy computation requirements. While DL methods currently cannot completely substitute conventional methods, they can speed up the 3D reconstruction speed given an initial guess, and can be fine-tuned using conventional methods to achieve better performances. For CDI 3D reconstruction, \cite{scheinker2020adaptive} proposes a 3D CNN architecture with model-independent adaptive feedback agents. The network takes in 3D diffracted intensities as inputs and outputs a vector of spherical harmonic coefficients, which describe the surface of the 3D incident object. The adaptive feedback agents take as input the spherical harmonics to adaptively adjust the intensities, positions, and decay rates of a collection of radial basis functions. The 3DCNN is trained using a synthetic dataset consisting of 500,000 training set of 49 sampling coefficients as well as the spherical surface and volume of each in order to perform a 3D Fourier transform. An additional 100 random 3D shapes and their corresponding 3D Fourier transforms are used to test the adaptive model-independent feedback algorithm, with the CNN output as its initial guess. Last, the robustness of the trained 3DCNN is tested on  the experimental data of a 3D grain from a polycrystalline copper sample measured using high-energy diffraction microscopy. Results show that the 3DCNN provides an initial guess that captures the average size and a rough estimate of the shape of the grain. The adaptive feedback algorithm uses the 3DCNN initial guess to fine-tune the harmonic coefficients to match and converge the generated and measured diffraction patterns of the grain.

\section{Hardware Solutions for Deep Learning}

Deep neural networks (DNNs) have been implemented for many imaging processing tasks ranging from enhancement to generation as discussed above. To achieve good performance, these algorithms use very deep networks which requires heavy computational power during training and inference. For very large networks such as AlexNet, a single forward pass may require millions of multiply and accumulate (MAC) operations, thus making DNNs both computationally and energy costly. For real-time data processing in imaging devices, DNN algorithms need to be executed with low latency, limited energy, and other design constraints. Hence, there is a need to develop cost and energy efficient hardware solutions for DL applications. 

Interestingly, neural network algorithms are known to have at least two types of inherent parallelism, namely model and data parallelism \cite{gholami2018integrated}. Model parallelism refers to the partitioning of the neural network weights for MAC operations for parallel execution as there are no data dependencies. Data parallelism refers to processing the data samples in batches rather than a single sample at a time. Hardware accelerators can exploit these characteristics by implementing parallel computing paradigms. This section presents different hardware accelerators used for DL applications. Note that the best hardware solution is dependent on the application and corresponding design requirements. For example, edge computing devices such as cameras and sensors may require small chip area with limited power consumption.

\begin{figure}[tb]
\centering
\includegraphics[width=0.8\linewidth]{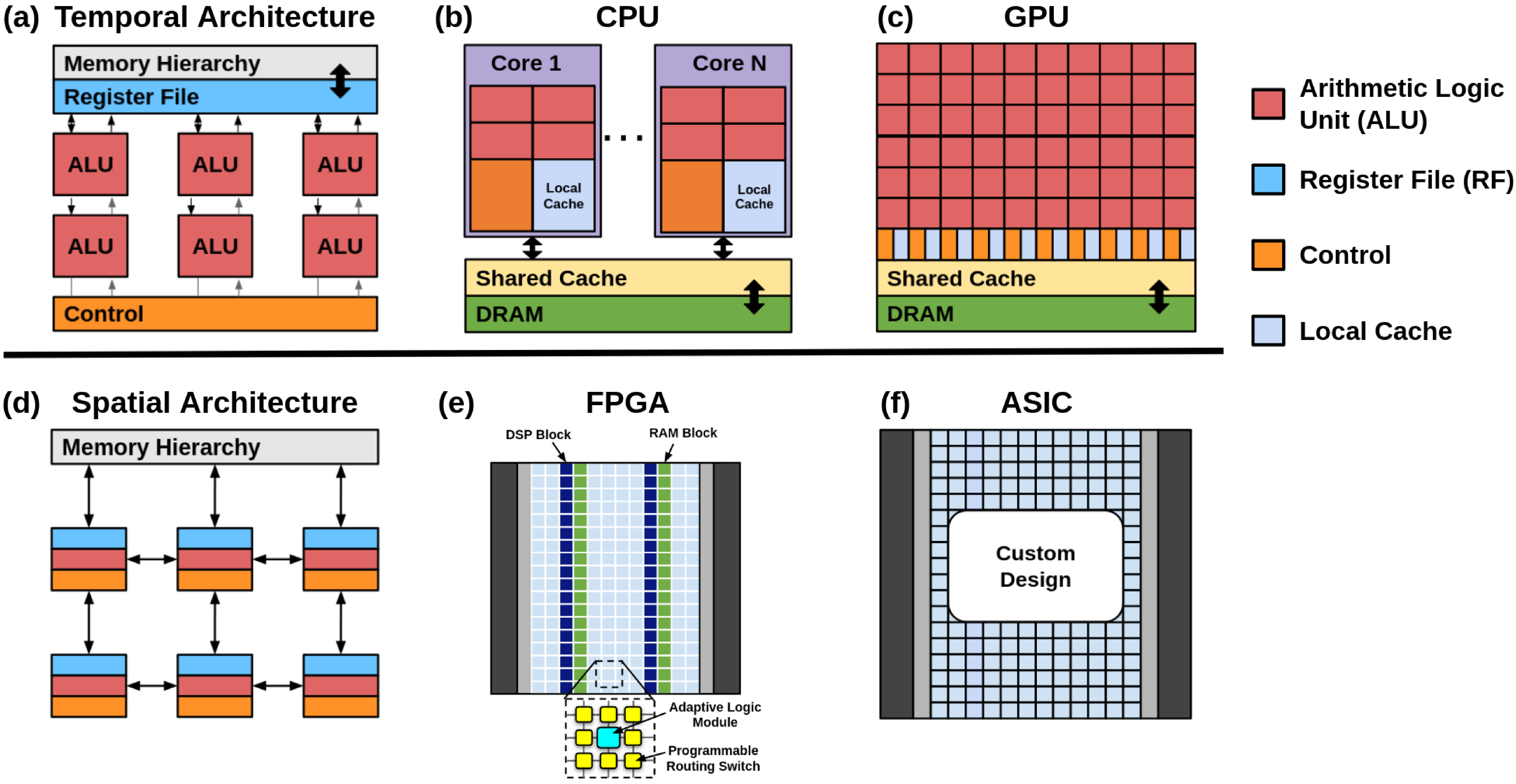}
\caption{Basic models of the (a) temporal, (b) CPU, (c) GPU, (d) spatial, (e) FPGA, and (f) ASIC architectures. }
\label{fig:architectures}
\end{figure}

\subsection{Electronic-based Accelerators}

The electronic-based hardware solutions for DL are broad, ranging from general purpose processors such as central processing units (CPUs) and graphical processing units (GPUs), field-programmable gate arrays (FPGAs), to application-specific integrated circuits (ASICs). The circuit architecture design typically follows either temporal or spatial architectures \cite{sze2017efficient} as shown in Figure \ref{fig:architectures}(a) and \ref{fig:architectures}(d). The architectures are similar in using multiple processing elements (PEs) for parallel computing, however, there are differences in control, memory, and communication. The temporal architecture features a centralized control for simple PEs, consisting of only arithmetic logic units (ALUs), which can only access data from the centralized memory. Meanwhile, the spatial architecture features a decentralized control scheme with complex PEs, where each unit can have its own local memory or register file (RF), ALU, and control logic. The decentralized control scheme forms interconnections between neighboring PEs to exchange data directly, allowing for dataflow processing techniques.

\subsubsection{Temporal Architectures: CPUs and GPUs}
CPUs and GPUs are general purpose processors that typically adopt the temporal architecture as shown in Figures \ref{fig:architectures}(b) and \ref{fig:architectures}(c). Modern CPUs can be realized as vector processors, which adopt the single-instruction multiple-data (SIMD) model to process a single instruction on multiple ALUs simultaneously. In addition, CPUs are optimized for instruction-level parallelism in order to accelerate the execution time of serial algorithms and programs. Meanwhile, modern GPUs adopt the single-instruction multiple threads (SIMT) model to process a single instruction across multiple threads or cores. Different from CPUs, GPUs are made up of more specialized, parallel, and smaller cores than CPUs to efficiently process vector data with high performance and reduced latency. As a result, GPU optimization relies on software defined parallelism rather than instruction-level parallelism \cite{intel2022}. Both the SIMD and SIMT execution models for CPUs and GPUs, respectively, allow for parallel MAC operations for accelerated computations.

Nonetheless, CPUs are not the most used processor for DNN training and inference. Compared to GPUs, CPUs have a limited number of cores, and thus a limited number of parallel executions. For example, one of Intel's server-grade CPUs is the Intel Xeon Platinum 8280 processor which can have up to 28 cores, 56 threads, 131.12 GB/s maximum memory bandwidth, and 2190 Giga-floating point operations per second (GFLOPS) for single-precision compute power. In comparison, NVIDIA's GeForce RTX 2080 Ti is a desktop-grade GPU with 4352 CUDA cores, 616.0 GB/s memory bandwidth, and 13450 GFLOPS single-precision compute power.

For DL at the edge, the hardware industry has developed embedded platforms for AI. One popular platform is the NVIDIA Jetson for next generation embedded computing. The Jetson processor features a heterogeneous CPU-GPU architecture \cite{mittal2015survey} where the CPU accelerates the serial instructions and the GPU accelerates the parallel neural network computation. Furthermore, the Jetson is designed with a small form factor, size, and power consumption. A broad survey by \cite{mittal2019survey} presents different works using the Jetson platform for DL applications such as medical, robotics, and speech recognition. Several surveyed works have used the Jetson platform to implement imaging processing tasks including segmentation, object detection, and classification. 

Also using the NVIDIA Jetson platform, a work by \cite{abeykoon2019scientific} investigates the performance of the Jetson TX2 for edge deployment for TomoGAN \cite{liu2020tomogan}, an image denoising technique using general adversarial networks (GANs) for low-dose X-ray images. The training and testing datasets consist of 1024 pairs of images of size $1024\times1024$ with each image pair consisting of a noisy image and its corresponding ground truth. The pre-trained TomoGAN network is deployed and tested on the Jetson TX2 and a laptop with an Intel Core i7-6700HQ CPU @2.60GHz with 32GB RAM. The laptop CPU achieves an average inference performance of 1.537 seconds per image, while the TX2 achieves an inference performance of 0.88 seconds per image, approximately $1.7\times$ faster than the laptop CPU.

A recent work by \cite{an2022tbnet} investigates the classification accuracy of tuberculosis detection from chest X-ray images using MobileNet \cite{sandler2018mobilenetv2}, ShuffleNet \cite{zhang2018shufflenet}, SqueezeNet \cite{iandola2016squeezenet}, and their proposed E-TBNet. In addition, they further investigate the inference time during testing of each network on the NVIDIA Jetson Xavier and a laptop with Intel Core i5-9600KF CPU and NVIDIA Titan V GPU. The dataset consists of 800 chest X-ray images scaled to size $512\times 512 \times 3$. The MobileNet network achieves the highest accuracy at 90\% while their proposed E-TBNet achieves 85\%. However, the inference time for E-TBNet is the fastest for all investigated networks with an inference time of 0.3 ms and 3 ms per image when deployed on the laptop with Titian GPU and Jetson Xavier, respectively. The slowest reported inference time for the Jetson Xavier is 6 ms per image for the ShuffleNet. Although the inference time for the Xavier is an order of magnitude slower, classification inference can be achieved in real-time with smaller hardware footprint for edge deployment.

\subsubsection{Spatial Architectures: FPGAs and ASICs}

Field-programmable gate arrays (FPGAs) and application-specific integrated circuits (ASICs) typically adopt the spatial architecture as shown in Figure \ref{fig:architectures}(e) and \ref{fig:architectures}(f). FPGAs and ASICs are specialized hardware that are tailored for specific applications due to their design process. FPGAs can be configured to perform any function as it is made up of programmable logic modules and interconnecting switches as shown in Figure \ref{fig:architectures}(e). The FPGA software is used to directly build the logic and data flow directly into the chip architecture. On the other hand, ASICs are designed and optimized for a single application, and cannot be reconfigured. Nonetheless, the spatial architecture of FPGAs and ASICs makes them well suited for neural network computations as the mathematical operations of each layer are fixed and known a priori. As a result, FPGAs and ASICs can attain highly optimized performance.

As shown in Figure \ref{fig:architectures}(d), the spatial architecture consists of an array of PEs interconnected with a Network-on-Chip (NoC) design, allowing for custom data flow schemes. Although not shown in Figure \ref{fig:architectures}(d), the memory hierarchy consists of three levels. The lowest level consists of the RF in each PE, which is used to locally store data for inter-PE data movement or local accumulation operations. The middle level consists of a global buffer (GB) that holds the neural network weights and inputs to feed the PEs. The highest level is the off-chip memory, usually a DRAM, to store the weights and activations of the whole network. MAC operations need to be performed on large data sets. Hence, the major bottleneck is the high latency and energy costs of DRAM accesses. A comparison between DianNao and Cambricon-X, two CNN accelerators, show that DRAM accesses consume more that 80\% of the total energy consumption \cite{li2018smartshuttle}. In addition, \cite{chen2016eyeriss} reports that the energy cost of DRAM access is approximately $200\times$ more than a RF access. Therefore, energy efficiency can be greatly improved through the reduction of DRAM accesses, commonly done by exploiting the idea of data reuse.

The focus of data reuse is to utilize the data already stored in RFs and the GB as often as possible. This gives rise to the investigations of efficient data flow paradigms in both the spatial and temporal operations of PEs. For example, in fully connected layers, the input reuse scheme is popular since the input vector is dot multiplied by each row of the weight matrix to compute the layer output. For convolutional layers, the weight reuse scheme is popular as the weight kernel matrix is used for multiple subsets of the input feature map. In addition for convolutional layers, convolutional reuse can be applied by exploiting the overlapping region of the sliding window of kernel weights and the input feature map. Additional data reuse schemes are the weight stationary, output stationary, row stationary, and no local reuse schemes. A detailed discussion of the data reuse schemes is out of scope for this paper. However, for a comprehensive review, see details in \cite{sze2017efficient, capra2020hardware, dhilleswararao2022efficient}. In summary, optimizing the data flow is crucial for FPGAs and ASICs to attain high energy efficiency.

The energy efficiency and massive parallelism of FPGA and ASIC-based accelerators make them desirable for edge computing. A recent work \cite{xia2021sparknoc} develops a lightweight CNN architecture called SparkNet for image classification tasks. SparkNet features approximately $3\times$ less parameters compared with the SqueezeNet, and approximately $150\times$ less parameters than AlexNet. In addition, a comprehensive design is presented to map all layers of the network onto an Intel Arria 10 GX1150 FPGA platform with each layer mapped to a its own hardware unit to achieve simultaneously pipelined work, increasing throughput. SparkNet is tested on 4 benchmark image classification datasets, i.e. MINIST, CIFAR-10, CIFAR-100 and SVHN. The performance and average time for the Intel FPGA, NVIDIA Titan X GPU, and Intel Xeon E5 CPU to process 10,000 $32\times 32 \times 3$ is reported. The FPGA-based accelerator achieves a processing time of 11.18 \textmu s, which is $41\times$ and $9\times$ faster than the CPU and GPU, respectively. Furthermore, the FPGA average power consumption is 7.58 W with a performance of 337.2 Giga operations per second (GOP/s), making the FPGA more energy and computationally efficient compared to the CPU (95 W, 8.2 GOP/s) and GPU (250 W, 39.4 GOP/s).

Another recent work \cite{liu2021collaborative} uses FGPAs to deploy MobileNet for face recognition in a video-based face tracking system. The work further integrates the FPGA with CPUs and GPUs to build a heterogeneous system with a delay-aware energy-efficient scheduling algorithm to achieve reduced execution time, latency, and energy cost. The face tracking experiment is run using an Intel Gold 5118 CPU, NVIDIA Tesla P100 GPU, and the Intel Arria 10 GX 900 and Intel Stratix 10 GX1100 FGPAs. The reported experimental results evaluate the computing speed and power efficiency of the FPGA-based accelerator compared to the CPU and GPU, as well as the efficiency of the combined detection system with CPU/GPU/FPGA. The FPGA accelerators achieve a computational speed that is approximate to or better than the GPU, while achieving superior power efficiency in GOP/s/W. The difference in performance of the FPGAs is due to their hardware specifications, where the hardware richer Intel Stratix will out perform the Intel Arria. Lastly, the experimental results report that the CPU/GPU/FPGA system can achieve optimal performance in comparison to using only one or a combination of two different accelerators. This is due to the energy efficient scheduling algorithm to optimally pipeline tasks to the different accelerators. The idea of utilizing a heterogeneous system and scheduling algorithm to improve computational and energy efficiency can be explored to address challenges of edge computing.

The Google Edge Tensor Processing Unit (TPU) platform \cite{cass2019taking} is a general purpose ASIC designed and built by Google for inference at the edge. One example product is the Dual Edge TPU which features an area footprint of $22 \times 30$ mm$^2$, peak perfrmance of 8 trillion operations per second (TOPS), and power consumption of 2 TOPS/W. Other hardware options are available for ASIC prototyping and deployment for edge devices. A survey by \cite{sun2021deep} presents works that use the Edge TPU platform for DL applications such as image classification, object detection, and image segmentation. 

The previously discussed work \cite{abeykoon2019scientific}, which deployed TomoGAN on the NVIDIA Jetson platform for X-ray image denoising, also deployed it on the Edge TPU. The work presents a quantized model of TomoGAN to address limitations of the Edge TPU, such as output size. A fine tuning model is also presented to improve the output quality of the quantized model. The Edge TPU's average inference time is 0.554 s per image, which is faster than the Jetson TX2 inference time of 0.88 s per image. In addition, the power consumption is reduced to 2 W compared to Jetson TX2's 7.5 W.

\begin{figure}[tb]
\centering 
\includegraphics[width = 0.3\linewidth]{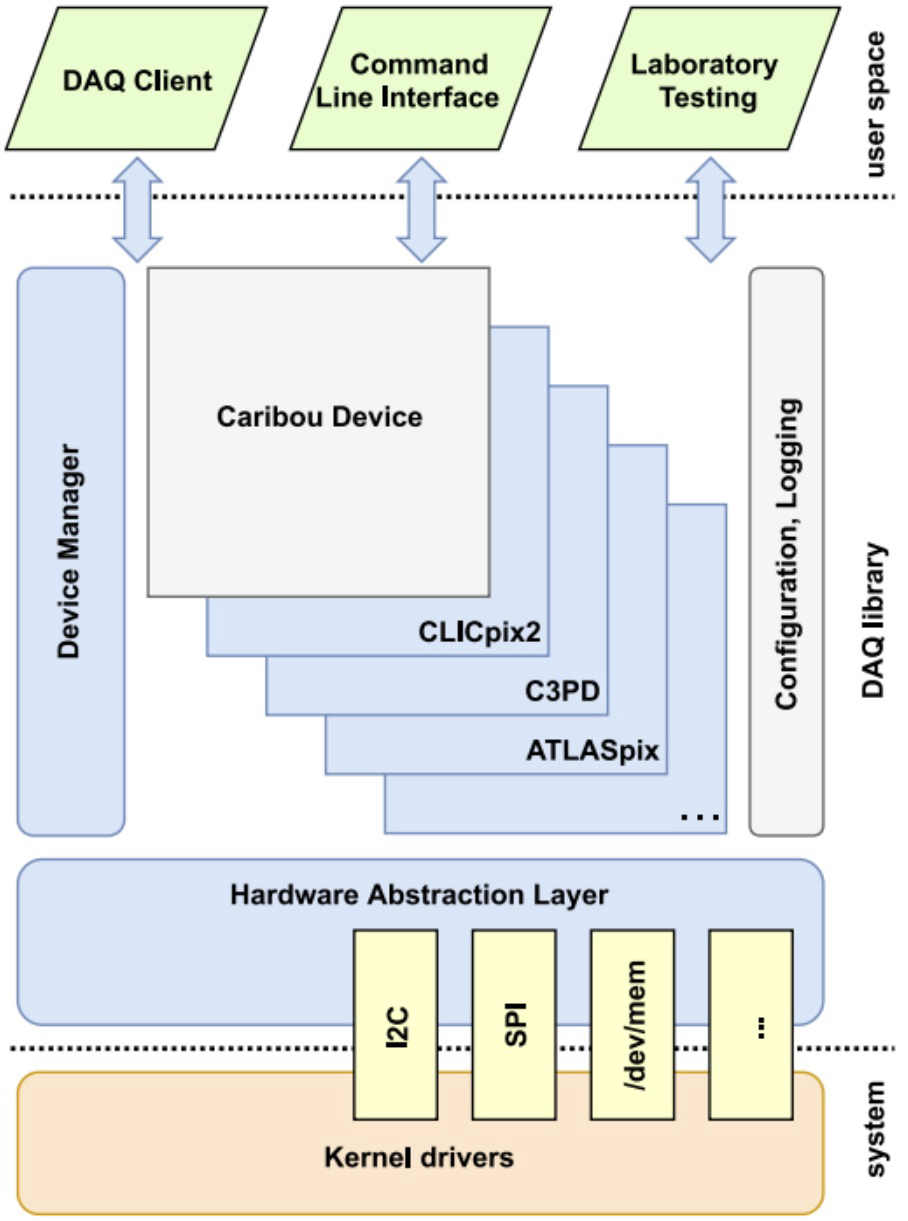}
\caption{Scalable CARIBOu architecture for data readout. Adopted from \cite{liu2017development} with permission.}
\label{fig:caribou}
\end{figure}

In addition, there is interest in the development of software and firmware for modular and scalable implementation of energy efficient algorithms on FPGA platforms. One such example is for high-speed readout systems for pixel detectors. Oak Ridge National Laboratory (ORNL), through the support of the Department of Energy (DOE) in High Energy Physics (HEP) and Nuclear Physics (NP), is leading the design of a new generic readout system for pixel detectors based on the successful first-generation system, the CARIBOu 2.0 \cite{liu2017development}. The CARIBOu 2.0 system, shown in Figure \ref{fig:caribou}, will be the proposed architecture for the platform. The concept of the system is to provide a generic framework for the readout of ASIC detectors for research and development and scalable to larger detector arrays. CARIBou 2.0 shares knowledge and code to provide the community with a convenient platform that maximizes reusability and minimizes overhead when developing such systems. ORNL will initially implement the readout firmware and software specific to the Timepix4 or to commercial CMOS image sensors, SMALLGAD, Photon-to-Digital Converters (PDCs), and the interconnect for the assemblies. The hardware platform is based on Xilinx Ultrascale+ FPGA, that provides resources for CPU and FPGA side data processing at high speed. 
Using the resources of this modern FPGA, software and firmware will be developed to flexibly implement data processing and reduction, edge computing, by using conventional and ML algorithms running in the FPGA. For larger data rates, firmware will be developed to move the data to a FELIX card, which can handle up to 24 CARIBOu 2.0 systems and transmit data via a high-speed network interface to a data center or process them locally via GPU and CPU in the FELIX host machine. As a result, the system can be scaled up to the readout of large smart sensor stack arrays.

Furthermore, as advancements in ASIC technology have enabled greater integration of digital functionalities for scientific applications, there has been growing interest in incorporating compression capabilities directly within ASIC detectors to enhance data processing speed. ASIC architectures capable of frame rates approaching 1 MHz have been designed, providing a viable solution for enhancing the speed of various diffraction techniques employed at X-ray light sources, including those relying on coherent imaging methodologies like ptychography \cite{strempfer2022lightweight}. Developing ASIC compression strategies that exploit the structure in detector data enables high compression performance while requiring lower computational complexity than commonly used lossless compression methods like LZ4 \cite{strempfer2022lightweight}.



\begin{figure}[tb]
\centering
\includegraphics[width=0.95\linewidth]{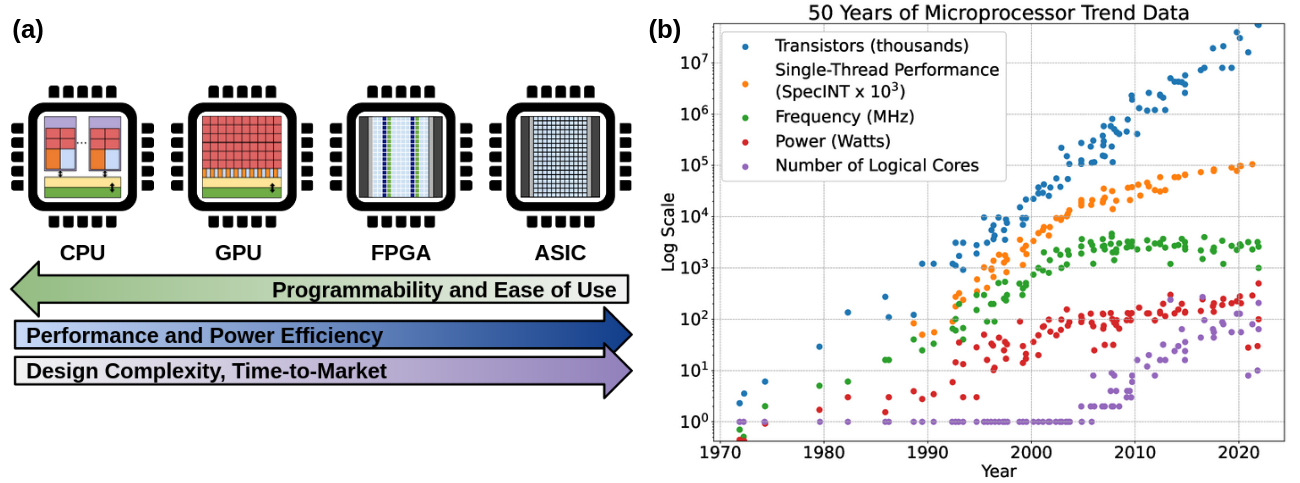}
\caption{(a) Summary of electronic-based hardware comparison. (b) 50 years of CPU processor trend data from \cite{rupp2022}.}
\label{fig:summary_trend}
\end{figure}

\subsubsection{Summary and Limitations}
We have presented an overview of 4 different electronic-based accelerators and a few works applying them to DL applications at the edge. Figure \ref{fig:summary_trend}(a) shows that there is a clear trade-off between programmability and efficiency. To attain higher performance and power efficiency, FPGAs and ASICs require more design complexity to optimize data flow, while ASICs need further hardware optimization. Correspondingly, the time-to-market increases with design complexity. DL algorithms can be deployed at the edge using these existing electronic-based hardware accelerators. However, in recent years, these electronic-based accelerators are constantly reaching performance limits in latency, energy consumption, high interconnect cost, excessive heat, and other physical constraints \cite{waldrop2016more}. Figure \ref{fig:summary_trend}(b) illustrates the past 50 years of CPU trends in regards to the number of transistors, single-thread performance, frequency, typical power consumption, and number of cores. The trends show that the number of transistors and correspondingly the power consumption continues to grow. As of November 2022, Apple's M1 Ultra chip has the largest number of transistors on a commercial processor at 114 billion. Furthermore, the trends indicate that CPU clock frequency has plateaued since around 2005 while single-thread performance and number of cores are slowly tapering. In addition, electronic accelerators are traditionally designed to follow von Neumann architecture where the processor and memory units are connected by buses \cite{heuring2021principles}, which inherently increases data transfer and power consumption during computation. \cite{li2018smartshuttle} demonstrates that more than 75\% of the energy utilized by processors comes from DRAM accesses. These limits in electronic based computing gives rise to a shift in focus to analog neuromorphic computing and non-von Neumann architectures such as optical neural networks and bio-inspired spiking nerual networks for high-speed, energy-efficient, and parallel computing \cite{ganguly2019towards,sui2020review}.

\subsection{Optical Neural Networks}

Optical neural networks (ONNs) have emerged as a promising avenue for achieving high-performance and energy-efficient computing, given their compute-in-light speed, ultra-high parallelism, and near-zero computation energy \cite{shen2017deep,shastri2021photonics,feng2023integrated,gu2022light}. Series of photonic tensor cores (PTCs) are designed to enhance the execution of linear matrix operations, the fundamental operations in AI and signal processing, with coherent photonic integrated circuits  \cite{shen2017deep}, micro-ring resonators \cite{tait2017neuromorphic}, photonic phase-change materials \cite{feldmann2021parallel,rios2019memory,zhu2022elight}, and diffractive optics \cite{lin2018all,yan2019fourier,ashtiani2022chip}. On the basis of the linear optical computing paradigms, ONNs have been constructed for various machine learning tasks such as image classification \cite{feng2022compact,feng2023optically,ashtiani2022chip}, vowel recognition \cite{shen2017deep}, and edge detection \cite{gu2020roq}. Photonic computing methods \cite{zhu2023dota} also feature great potential for supporting advanced Transformer models. Furthermore, ONNs holds significant promise for real-time image processing, where they process image signals directly in light fields, as opposed to after digitalization \cite{wang2023image,zhou2023ultrafast,yamaguchi2023time,huang2023photonic}. For instance, recent advancements include the proposal of an image sensor with an ONN encoder \cite{wang2023image}, which filters relevant information within a scene using an energy-efficient ONN decoder before detection by image sensors.

Despite their advantages, PTCs face significant challenges related to cross-domain signal conversion energy overhead, specifically in analog-to-digital (A/D) and digital-to-analog (D/A) conversion, as well as scalability, particularly concerning chip area. For example, Mach-Zehnder interferometer (MZI)-based PTCs \cite{shen2017deep} require $O(m^2 + n^2)$ bulky MZIs and approximately $\sim (m+n)$ cascaded MZIs within a single optical path to implement an n-input, m-output layer.

Efficient analog-to-digital conversion solution \cite{zhu2022fuse} and various hardware-software co-design methodologies \cite{gu2020roq} have been investigated to reduce signal conversion overhead by reducing precision and energy per conversion.

\begin{figure}[tb]
\centering
\includegraphics[width=0.6\linewidth]{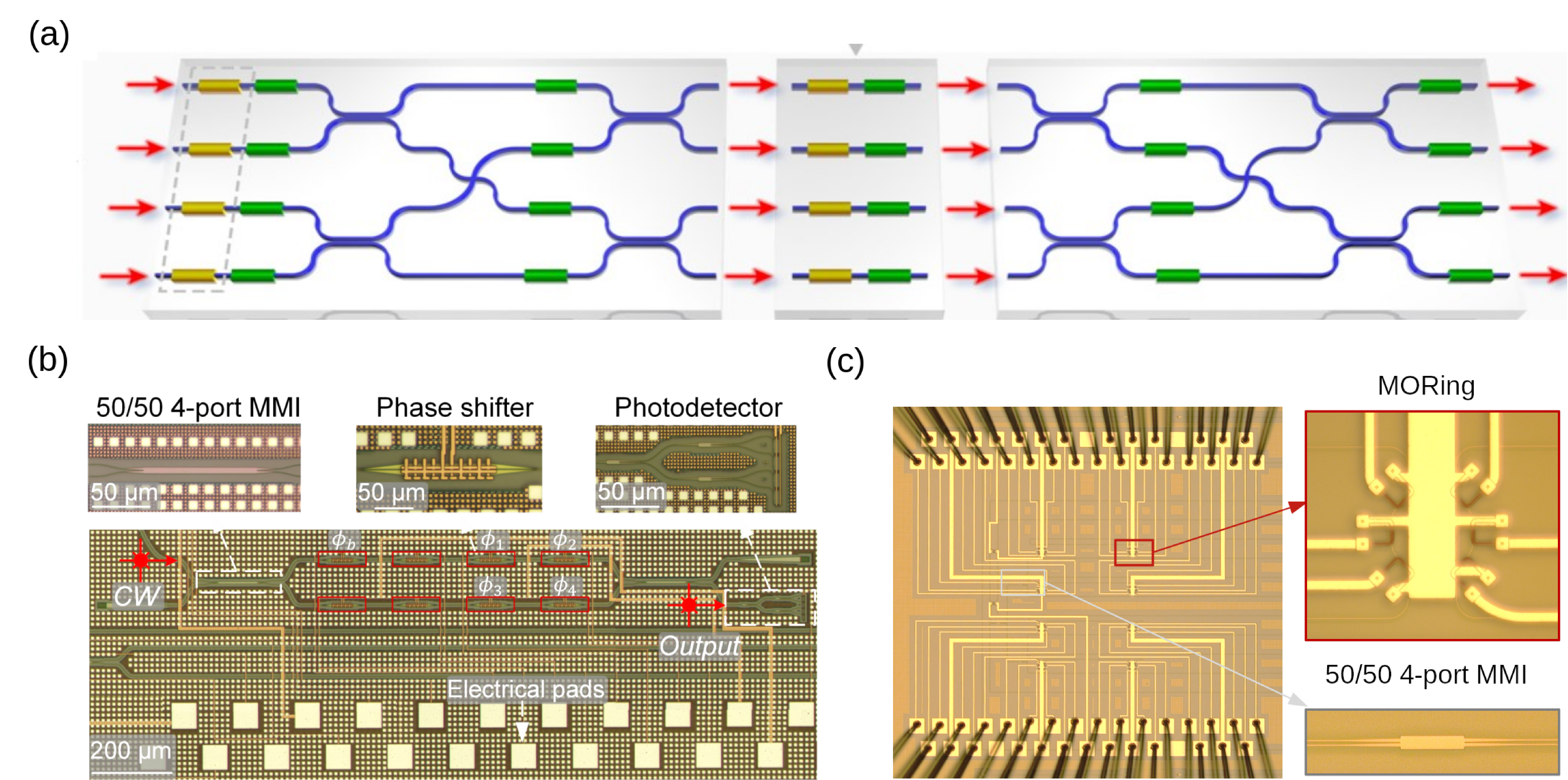}
\caption{Integrated photonic chips for optical neural networks. (a) a butterfly-style PTCs to reduce the opitcal components from an architecture level \cite{feng2022compact}. (b) and (c) are customized multi-operand MZI-based and microring resonator-based PTCs, respectively, which improve scalability and efficiency at the device level \cite{gu2022squeezelight,feng2023integrated2,feng2023optically}.}
\label{fig:onn}
\end{figure}

In pursuit of enhancing the scalability and efficiency of ONNs, researchers have delved into innovative optimizations at both the architecture and device levels. One noteworthy approach at the architecture level is the introduction of optical subspace neural networks (OSNNs), which make a trade-off between weight representation universality and the reduction of optical component usage, area costs, and energy consumption. For example, a butterfly-style OSNN as shown in Figure \ref{fig:onn}(a), which achieved a remarkable reduction of 7 times in trainable optical components compared to GEMM-based ONNs, was reported and demonstrated a measured accuracy of 94.16\% in image recognition tasks \cite{feng2022compact}. Without sacrificing much model expressiveness, OSNNs can reduce footprints, often ranging from one to several orders of magnitude less than previous MZI-based ONN \cite{shen2017deep}.

At the device level, employing compact custom-designed PTCs, such as multi-operand optical neurons (MOON) \cite{gu2022squeezelight,feng2023integrated2,feng2023optically}, enables the consolidation of matrix operations into arrays of optical components. Figure \ref{fig:onn}(b) and \ref{fig:onn}(c) shows a customized multi-operand MZI-based and microring resonator-based PTCs, respectively. Instead of performing a single math operation (e.g., scalar product) per device, MOON fuses a tensor operation in the single device. Crucially, this approach retains the capability to represent general matrices while still maintaining an exceptionally compact layout, in contrast to prior compact tensor designs like star couplers and metasurfaces \cite{zhu2022space}. One specific achievement in MOON is the development of multi-operand MZI-based (MOMZI) ONN \cite{feng2023integrated2}, which has realized a two-orders-of-magnitude reduction in propagation loss, delay, and total footprint without losing matrix expressivity. The customized ONN demonstrated an 85.89\% measured accuracy in the street view house number (SVHN) recognition dataset with 4-bit control precision. The combined progress in architecture, device design, and optimization techniques is pivotal in advancing the capabilities of ONNs, making themselves efficient, scalable, and practical for AI applications.

\subsection{Spiking Neural Networks}

\begin{figure}
\centering
\includegraphics[width=0.75\linewidth]{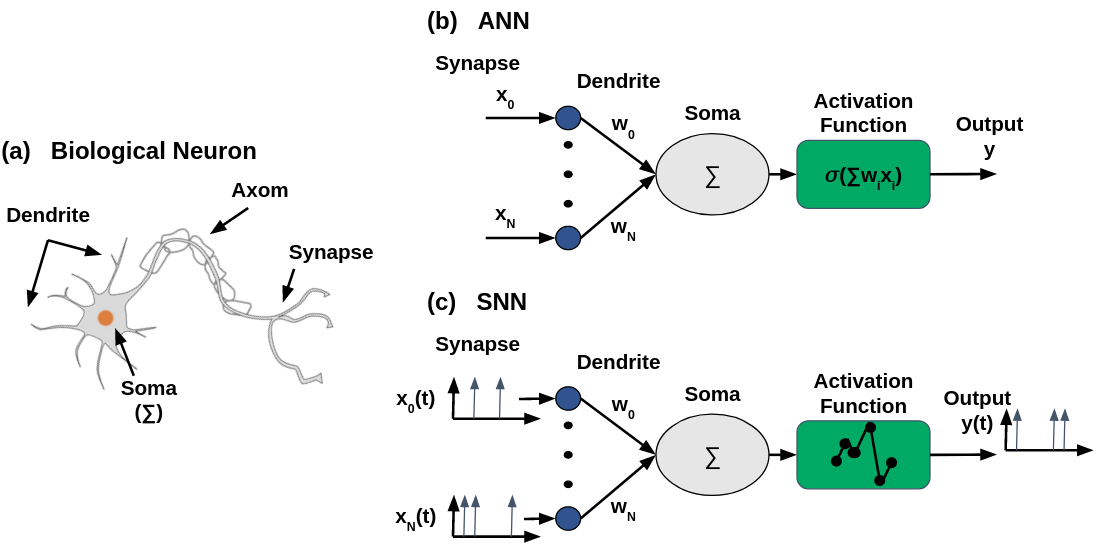}
\caption{A comparison among (a) a biological neuron, (b) artificial network neuron, and (c) spiking network neuron.}
\label{fig:snn}
\end{figure}

In addition to photonic neuromorphic computing, extensive research has been done for other neuromorphic computing architectures. Due to the bottleneck seen in von Neumann architectures, these computing paradigms aim to greatly reduce data movement between memory and PEs to attain high energy efficiency and parallel processing. Taking a unique approach to improve energy efficiency, neuromorphic computing architectures are inspired by the human brain's neurons and synapses. The human brain is extremely energy efficient, where in terms of computing terminology, it is estimated to have a computing power of 1 exaFLOPS while only consuming 20 W. In recent years, there is a rise in interest to explore brain-inspired neural network computing architectures, better known as spiking neural networks (SNNs) \cite{nunes2022spiking}. 

SNNs are a special type of artificial neural network (ANN) that closely mimics biological neural networks. While ANNs are traditionally modeled after the brain, there are still many fundamental differences between them such as neuron computation and learning rules. In addition, one major difference is the propagation of information between neurons. Biological neurons, shown in Figure \ref{fig:snn}(a), transmit information to downstream neurons using a spike train of signals, or a time-series of delta functions. The individual spikes (delta functions) are known to be sparse in time and have high information content. Therefore, SNNs are designed to convey information by utilizing the spike timings and spike rates \cite{tavanaei2019deep, schuman2022opportunities} as shown in Figure \ref{fig:snn}(c). Furthermore, the advantages of the spiking event sparsity can be exploited in special hardware to reduce energy consumption while maintaining the transmission of high information content \cite{pfeiffer2018deep}. 

The hardware industry as well as academia are striving to develop unique solutions for neuromorphic computing chips. Intel's Loihi \cite{davies2018loihi} features 128 neuromorphic cores with 1024 spiking neural units per core. A recent work \cite{davies2021advancing} surveys different works that utilize Loihi as a computing platform for applications such as event-based sensing and perception, odor recognition, closed-loop control for robotics, and simultaneous localization and mapping. For medical image analysis, \cite{getty2021deep} uses Loihi to implement a SNN for brain cancer MRI image classification. IBM developed TrueNorth \cite{merolla2014million}, a neurmorphic chip featuring 4096 neuromorphic cores, 1 million spiking neurons and 256 million synapses. A work by \cite{shukla2019remodel} uses the TrueNorth computing platform to detect and count cars from input images by mapping CNNs, such as AlexNet and VGG-16, onto TrueNorth. A few other well-known SNN hardwares are Neurogrid \cite{benjamin2014neurogrid}, BrainScaleS \cite{schemmel2010wafer}, and SpiNNaker \cite{furber2014spinnaker}, which all adopt different solutions to emulate spiking neurons. For a comprehensive review, see details in \cite{bouvier2019spiking, basu2022spiking}.

Due to its low power consumption, SNN hardware is a potential platform for edge computing. A work by \cite{schuman2017neuromorphic} presents preliminary results for implementing SNN on a mixed analog digital memresistive hardware for classifying neutrino scattering data collected at Fermi National Accelerator Laboratory using the MINERvA detector \cite{aliaga2014design}. Two different SNNs, the neuroscience-inspired dynamic architecture (NIDA) \cite{schuman2014spatiotemporal} and a memresistive dynamic adaptive neural network array (mrDANNA) \cite{cady2019development}, were trained and tested on the MINERvA dataset's X view. The training and testing datasets consisted of 10,000 and 90,000 synthetic instances, respectively, generated by a Monte Carlo generator. The NIDA network was trained on the Oak Ridge Leadership Computing Facility's Titan using 10,000 computing nodes, and achieved a classification accuracy of 79.11\% on the training set. Meanwhile, the mrDANNA was trained on a desktop and achieved a classification accuracy of 76.14\% and 73.59\% on the training and combined training and testing dataset, respectively. Both networks can attain an accuracy close to the state-of-art CNN accuracy of 80.42\% while using far less neurons and synapses. In addition, the energy consumption was computed for the mrDANNA network and is estimated to be 1.66 $\mu$J per calculation. Although there is an accuracy drop using the smaller SNN networks, the energy consumption per calculation is very small, and thus can  be deployed in edge devices.

A recent work \cite{r2023sensor} implemented an SNN algorithm for filtering data from edge electronics in high energy collider experiments conducted at the High Luminosity Large Hadron Collider (HL-LHC), in order to reduce large data transfer rate or bandwidth (on the order of a few petabytes per second) to downstream electronics. In collider experiments, the collision events of charged particles with energy greater than 2 GeV is of significant interest. However, the high energy charged particles only comprise of approximately 10\% of all recorded collision events. Therefore, filtering out low energy particle track clusters will greatly reduce data collection rate at edge devices. A synthetic dataset is used to train and test the SNN. The full synthetic dataset consists of 4 million charged particle interactions in a silicon pixel sensor. The training dataset is limited to the particle interactions in a $13\times 21$ pixel sub-region of the silicon sensor, with binary classification labels indicating high or low energy. The SNN is realized on Caspian \cite{mitchell2020caspian}, a neuromorphic development platform, and achieved a signal classification accuracy of 91.89\%, very close to a prototyped full-precision DNN accuracy of 94.8\%. In addition to accuracy, the SNN achieves good performance using nearly half of the number of DNN parameters. The reduced size and improved power efficiency of the SNN model makes it a good candidate for deployment on edge devices which have limited memory and power constraints.

\section{Summary}

Experimental data generation at photon sources are rapidly increasing due to the advancements in light sources, detectors, and more efficient methods or modalities to collect data. As tabulated in Table \ref{tab1:data}, detectors can achieve frame-rates on the order of millions of frames per second with at least 10-bit data resolution, and thus can achieve data rates over 1 Gb/s in continuous mode, and orders of magnitude higher data rate in burst mode. The high data rate is very costly in terms of data storage and transmission over long distances. These issues motivates the use of edge computing on detectors for real-time data processing and for reducing data transmission latency and storage volumes. 

Deep learning approaches have achieved significant progress in image processing tasks including but not limited to restoration, segmentation, compression, and 3D reconstruction. Their superior nonlinear approximation capabilities allow them to learn complex underlying structures and patterns in high dimensional data. The state-of-the-art methods for each image processing task achieve superior performance compared to conventional methods, while also overcoming the issues of conventional methods such as computational burdens associated with explicit programming for each data processing steps. Furthermore, once trained, deep learning methods can achieve very fast inference speeds for real-time computation. 

While deep learning approaches are widely used for many applications, they require deep networks to achieve good performance, and thus require heavy computational power and high energy consumption. This is critical hurdle for edge computing devices which have design constraints such as latency and energy. To address this issue, hardware accelerators now exist that leverage the model and data parallelism characteristics of neural network algorithms to implement parallel computing paradigms. Electronic-based hardware accelerators such as CPUs, GPUs, FPGAs, and ASICs are popularly used platforms for deep learning. However, the electronic-based solutions are constantly reaching performance limitations in clock speed, energy consumption, and other physical constraints. This gives rise to research in analog neuromorphic computing paradigms such as ONNs and SNNs to achieve high-speed, energy-efficient, and high-parallel computing, with significant potential for radiation detection and applications in photon science.

\section*{Conflict of Interest Statement}

The authors declare that the research was conducted in the absence of any commercial or financial relationships that could be construed as a potential conflict of interest.

\section*{Author Contributions}

SL: Supervision, Writing - original draft, Writing - review \& editing; SN: Writing - original draft, Writing - review \& editing; HZ: Writing - original draft, Writing - review \& editing; TZ: Writing - original draft, Writing - review \& editing; CLM:	Writing - review \& editing; SC: Writing - review \& editing; MC: Writing - original draft, Writing - review \& editing; RTC: Conceptualization, Writing - review \& editing; ZW: Conceptualization, Writing - original draft, Writing - review \& editing


\section*{Funding}

LANL work was performed under the auspices of the U.S. Department of Energy (DOE) by Triad National Security, LLC, operator of the Los Alamos National Laboratory under Contract No. 89233218CNA000001, including LANL Laboratory Directed Research and Development (LDRD) Program. This work is also supported in part by AFOSR MURI research center on Energy-efficient Optical Interconnects and Computing (Cont. No. FA9550-17-1-0071 managed by Dr. Gernot Pomrenke).

\section*{Acknowledgments}

SL and ZW wish to thank Dr. Alice Bean from University of Kansas for reviewing the hardware section. SL and ZW wish to thank Dr. Mathieu Benoit from Oak Ridge National Laboratory for the CARIBOu discussions.

\newpage

\bibliographystyle{ieeetr}
\bibliography{ref}

\end{document}